\documentclass{emulateapj}
\usepackage{amsmath}
\usepackage{setspace} 
\usepackage{graphicx}
\usepackage{color}
\usepackage[english]{babel}
\usepackage[utf8]{inputenc} 
\usepackage{slashed}
\usepackage{hyperref}
\usepackage{epstopdf}
\usepackage{natbib}
\usepackage{hhline}
\usepackage[toc]{appendix}
\begin{document}

\title{AGN evolution from a galaxy evolution viewpoint}
\author{Neven Caplar}
 \email{neven.caplar@phys.ethz.ch}
\author{Simon J. Lilly}
\author{Benny Trakhtenbrot\altaffilmark{1}}
\altaffiltext{1}{Zwicky Fellow}
\affiliation{%
Institute for Astronomy, Department of Physics, ETH Zurich, Wolfgang-Pauli-Strasse 27, 8093, Zurich, Switzerland
}%

  \today

\begin{abstract}
We explore the connections between the evolving galaxy and AGN populations. We present a simple phenomenological model that links the evolving galaxy mass function and the evolving quasar luminosity function, which makes specific and testable predictions for the distribution of host galaxy masses for AGN of different luminosities. We show that the $\phi^{*}$ normalisations of the galaxy mass function and of the AGN luminosity function closely track each other over a wide range of redshifts, implying a constant ``duty cycle" of AGN activity. 
The strong redshift evolution in the AGN  $L^*$ can be produced by either an evolution in the distribution of Eddington ratios, or in the $m_{bh}/m_{*}$ mass ratio, or both. To try to break this degeneracy we look at the distribution of AGN in the SDSS ($m_{bh},L$) plane, showing that an evolving ratio $m_{bh}/m_{*}  \propto (1+z)^2$ reproduces the observed data and also reproduces the local relations which connect the black hole population with the host galaxies for both quenched and star-forming populations.  We stress that observational studies that compare the masses of black holes in active galaxies at high redshift with those in quiescent galaxies locally will always see much weaker evolution. Evolution of this form would produce, or could be produced by, a redshift-independent $m_{bh} - \sigma$ relation and could explain why the local $m_{bh} - \sigma$ relation is tighter than $m_{bh} - m_{*}$ even if $\sigma$ is not directly linked to black hole growth.
Irrespective of the evolution of $m_{bh}/m_{*}$, the model reproduces both  the appearance of ``downsizing'' and the so-called ``sub-Eddington boundary'' without any mass-dependence in the evolution of black hole growth rates.

\end{abstract}


\maketitle

\section{Introduction}

Over the last few years there has been a lot of interest in the cosmic ``co-evolution'' of supermassive black holes (SMBH) and the stellar populations of the galaxies that they reside in.  This has been motivated on the one hand by the tight scaling relations that have been established between the masses of the black holes and various parameters  of the host galaxies  (e.g. \citealp{Kor95}; \citealp{Mag98}; \citealp{Fer00}; \citealp{Geb00}; \citealp{Tre02}; \citealp{Mar03}; \citealp{Har04}; \citealp{Fer05}; \citealp{Gre06}; \citealp{Gre07}; \citealp{All07}; \citealp{Gre08}; \citealp{Gra11}; \citealp{San11}; \citealp{Vik12}; \citealp{Mcc13})
and on the other by the overall similarities between the evolution of the star-formation rate density of the Universe and the luminosity density that is ascribed to SMBH accretion (e.g. \citealp{Boy98}; \citealp{Fra99}; \citealp{Mar04}; \citealp{Hec04}; \citealp{Has05}; \citealp{Sil09}; \citealp{Zhe09}). This paper seeks to link directly the evolution of the galaxy and AGN populations via a simple phenomenological model.\\
  
Looking first at galaxies, almost all galaxies can be broadly classified into a few distinct populations. The majority of star-forming (SF) galaxies have a star-formation rate (SFR) that is strongly correlated to their existing stellar mass, producing the so-called ``Main Sequence'' in which the specific SFR (or sSFR) varies only weakly with stellar mass.  This characteristic sSFR of the Main Sequence however increases strongly with look-back time (\citealp{Dad07}; \citealp{Elb07}; \citealp{Pan09}; \citealp{Spe14}) and is about a factor of twenty higher at $z \sim 2$ compared with the present day value. A small percentage of star-forming galaxies have significantly elevated sSFR, above the Main Sequence. 
It appears that the fraction of these ``outliers'' is more or less constant to z $\sim$ 2 \citep{Sar12} and that they represent of order 10\% of the integrated star-formation that is occurring at any epoch.   There is also a large population of ``quenched'' galaxies in which the star formation is substantially lower than on the Main Sequence, producing an sSFR that is much lower than the inverse Hubble time. Our understanding of the physical processes that lead to the quenching of star-forming galaxies is still quite limited and a number of plausible physical mechanisms have been proposed (some of which involve AGN directly).  However, the main empirical, or phenomenological, features of this quenching process are quite well understood based on the characteristic of the evolving population(s) of galaxies. \\

 As the available data on the galaxy population at substantial look-back times has improved, new analysis techniques have been introduced that take a purely empirical (phenomenological) approach to the data, see e.g \cite{Pen10} and \cite{Beh13}. These are complementary to the semi-analytic models of galaxy evolution (e.g. \citealp{Som08}, \citealp{Con09}, \citealp{Hen14}). \\
 
 The approach in \cite{Pen10} was to identify a few striking simplicities exhibited by the galaxy population(s) and to explore the consequences of these, where possible analytically, in terms of the most basic continuity equations linking the galaxy population(s) at different epochs. The \cite{Pen10} analysis was based on dividing the galaxy population into two components, the star-forming Main Sequence (including the outliers) and the quenched population of passive galaxies.  Much of the  \cite{Pen10} formalism is based on the observation that the characteristic Schechter $M^{*}$ of the mass-function $\phi(m)$ of the star-forming population has been more for less constant back to at least $z \sim 2.5$ (and likely to $z \sim 4$) despite the substantial increase in stellar mass (by a factor of 10-30) of any galaxy that stays on the star-forming Main Sequence over this same time period (\citealp{Bel03}; \citealp{Bel07}; \citealp{Ilb10}; \citealp{Poz10}; \citealp{Ilb13}). The \cite{Pen10} continuity formalism is very successful at reproducing the single and double Schechter \citep{Pre74} shapes of the mass functions of star-forming and passive galaxies in SDSS and also explains the quantitative relations between the Schechter parameters of these mass functions which constitute a test of the approach.  The formalism allows easy computation of things like the quenching rate of galaxies, the mass function of galaxies that are undergoing quenching at any epoch, and so on. The alternative phenomenological approach of  ``abundance matching" (\citealp{Beh13}) provides similar results, in terms of mass functions and star formation histories. With these recent developments, we now have a self-consistent,
empirical ("phenomenological") description of the evolving galaxy population at least back to z $\sim$ 4 in terms of the evolving mass-functions of both the star-forming and quenched populations of galaxies. \\
 
Turning to the AGN, the most basic description of the evolving population is the bolometric luminosity function (i.e. quasar luminosity function, QLF), i.e. $\phi(L,z)$. Large homogeneous samples of AGN have been created, from optical surveys such as Sloan Digital Sky Survey (SDSS) (\citealp{Schn10}; \citealp{Ros13}) and deep X-ray surveys, out to redshifts $z \sim 5$ (e.g.  \citealp{Has05}; \citealp{Bro06}; \citealp{Sil08}; \citealp{Bru10}; \citealp{Civ11}; \citealp{Kal14}; \citealp{Ued14}). These QLF studies have found that the QLF $\phi(L)$ is best described by a double power law, i.e. two power-law segments broken at a characteristic luminosity $L^{*}$,
\begin{equation} \label{eq:QLFF}
\phi(L)\equiv \frac{d N}{d \mbox{log} L} = \frac{\phi^{*}_{QLF}}{(L/L^{*})^{\gamma_{1}} +(L/L^{*})^{\gamma_{2}}}.
\end{equation}
These studies tend to agree that the increase in the number density of luminous quasars with increasing redshift is mostly driven by an evolution of $L^{*}$ at redshifts below $z \sim 2$. The exact shape of the QLF and need for evolution of other parameters is still debated (\citealp{Ued03}; \citealp{Bar05}; \citealp{Cro09}; \citealp{Air10}; \citealp{Ass11}). \\

The double power-law shape of the QLF contrasts the galaxy mass function $\phi(m)$ which is clearly better described by one or more Schechter functions, i.e. a power-law at low masses (or luminosities) and an exponential cut-off at masses above a characteristic mass $M^*$.  This difference in the shapes of these two most basic descriptions of the two po\interfootnotelinepenalty=10000pulations is interesting in that AGN are being harboured in galaxies and one might naively expect similarities between these two functions. This difference will form an important part of the current analysis.\\

The luminosity of an individual AGN can be expressed as 
\begin{equation} \label{eq:LumWithEd}
L =10^{38.1} \cdot  \lambda  \cdot m_{bh}
\end{equation}
where $L$ is given in erg s$^{-1}$, $\lambda$ is the Eddington ratio and $m_{bh}$ is the mass of the central black hole in units of solar mass. The SMBH mass is therefore a key quantity in understanding both an individual AGN and the AGN population. Unfortunately, black hole masses can be measured reliably with dynamical modeling only for small number of nearby sources in which the SMBH is generally quiescent (e.g. \citealp{Gul09}; \citealp{Schu11}; \citealp{Gra13}; \citealp{Mcc13}; \citealp{Rus13}; \citealp{Kor13} and references within).  Objects that are actively accreting and/or that are further away need to be analyzed with different techniques. For AGNs that show broad lines in their spectra it is possible to determine $m_{bh}$ by conducting reverberation mapping campaigns (see review by \citealp{Pet13} and references within). The results of such campaigns makes it possible to construct ``single-epoch" mass estimators, which allow the determination of $m_{bh}$ of AGN based on luminosity and line width. Systematic uncertainties in these estimators are calibrated using a sample of local AGNs for which black hole masses are statistically known from the $m_{bh}-\sigma$ of local (inactive) galaxies.   \\

Early suggestions that the distribution of specific accretion rates onto black holes has no dependence on black hole or galaxy mass (\citealp{Yu05}; \citealp{Kol06}; \citealp{Mer08}), have gained further observational support from analysis of the PRIMUS and COSMOS survey fields (\citealp{Air12}, \citealp{Bon12}). We will adopt a similar assumption in this paper. \\

Observations in the local Universe have revealed several interesting correlations between the mass of the SMBH in the center of a given galaxy and various quantities describing the surrounding galaxy.  Specifically, quite tight relations have been established with the stellar velocity dispersion ($m_{bh}-\sigma$) and with the stellar mass of the bulge in the galaxy ($m_{bh}-m_{bulge}$) (\citealp{Mag98}; \citealp{Fer00}; \citealp{Geb00}; \citealp{Mar03}). Canonical values of the ratio between $m_{bh}$ and $m_{bulge}$ are between $10^{-3}$ and $10^{-2.7}$  \citep{Har04}. The recent extensive review by \cite{Kor13} has argued for larger values, up to $10^{-2.3}$, mainly due to differentiation between bulges and pseudo-bulges, by omitting mergers in progress and by the inclusion of dark matter into dynamical modelling. 
Other scaling relations have been proposed, such as $m_{bh}-n$ where $n$ is Sersic index describing the light profile of the galaxy, $m_{bh}-m_{halo}$ where $m_{halo}$ is mass of the dark matter halo and $m_{bh}-m_{*}$ where $m_{*}$ is integrated stellar mass of the galaxy \citep{San11}. All these relations tend to show more scatter, but the basic difficulty is that most of these galaxy parameters are strongly correlated with each other.  It has been claimed that the correlation between $m_{bh}-m_{*}$ is more fundamental then $m_{bh}-m_{bulge}$ relation \citep{Mar13}, especially in star-forming hosts, which could then help to explain the AGN sources in bulgeless, pure disk, galaxies that have been observed \citep{Sim13}. In this paper we will base the analysis on $m_{bh}-m_{*}$ simply because the available mass functions of galaxies generally utilise the integrated stellar mass rather than the bulge mass, especially at high redshift. \\

In fact, almost all of the properties of galaxies that host AGN are still widely debated. Although there are undoubtedly some active AGN found in quenched galaxies, the bulk of radiatively efficient AGN seem to reside in star-forming galaxies (\citealp{Net09}; \citealp{Sil09}; \citealp{Sch10}; \citealp{Kos11}; \citealp{Cim13}; \citealp{Rosa13}; \citealp{Mat14}), especially in those relatively low luminosity systems where the host galaxy can be most easily discerned. Nevertheless, it is still not entirely clear what fraction of AGN are situated in quenched galaxies and whether this fraction changes with AGN luminosity.\\

Many studies have suggested that there is some redshift evolution in mass scaling relations (\citealp{Pen06}; \citealp{Dec10}; \citealp{Mer10}; \citealp{Tra10}; \citealp{Ben11}; \citealp{Sij14}), but others have found no significant evolution (\citealp{Jah09}; \citealp{Cis11}; \citealp{Mul12}; \citealp{Schr13}) or have argued that an observed evolution can be fully explained with selection effects (\citealp{Schu11}; \citealp{Schu14}). The redshift evolution of the scaling relations is still debated and we will explore different possibilities in the current paper within our model framework. We will also emphsize the methodological issues associated with the choice of samples.\\

Several studies have also attempted to constrain different aspects of black hole evolution with empirical or semi-empirical approaches which model the evolving black hole mass function, accretion rate, QLF, duty cycle,  accretion efficiency or some subset of these quantities (\citealp{Mer04}; \citealp{Sha04}; \citealp{Yu04}; \citealp{Mer08};  \citealp{Hop09}; \citealp{Sha09}; \citealp{She09}; \citealp{Cao10}; \citealp{Li11}; \citealp{Ste11d}; \citealp{Sha13}; \citealp{Con13}; \citealp{Nov13}; \citealp{Gou14}).\\

 In this work, we aim to apply a similar phenomenological approach that has proved so successful in describing the evolving galaxy population evolution to the evolving AGN population. We will take the simplest observables of the population and try to infer the underlying simplicities of the situation and use straightforward prescriptions to construct a simple model that can both explain the salient observational data and which can be used to make simple testable predictions for other quantities.  A focus of this work is to try to use our improved understanding of the evolving galaxy population, and especially of the evolving mass function $\phi(m_{*})$ of galaxies to high redshifts, to interpret the evolution of the AGN population. We aim thereby to create a simple global model to interpret the evolving AGN population and to enable us to evaluate biasses that may arise in observational work, e.g. through the use of luminosity-selected samples. The model is based on the construction of the AGN QLF via a double convolution of the underlying host galaxy mass function with a scatter function representing the black-hole to stellar mass ratio, and with an Eddington ratio distribution. \\ 

\cite{Hop081} and \cite{Hop10} have also used the observed galaxy mass functions as a basis for modelling the AGN population using AGN lightcurves obtained from simulations. A more similar approach to ours has been used by \cite{Air13} with an observationally driven model that connects AGN and galaxy evolution out to z $\sim$ 1, based on observational data from the PRIMUS survey \citep{Air12}. They have shown that is possible to reproduce the luminosity function measurements with a mass-independent distribution of specific accretion rates, i.e. linking the AGN QLF to the galaxy {\it mass function}.   \cite{Hic14} has also used a simple phenomenological approach and shown that the main observational trends (such as QLF and SFR-$L_{AGN}$ correlations) can be recovered by assuming that all star-forming galaxies host an AGN and that star formation and black hole accretion are correlated on $\sim$ 100 Myr time-scales, i.e. in this case they were linking the {\it growth} of the black holes and stellar populations.   A very recent paper of \cite{Vea14} has, like the current work, explored the links between the evolving galaxy (and halo) mass functions and the observed AGN QLF via a convolution approach.  \cite{Vea14} have emphasised the degeneracies between the ``mass function'' and ``mass growth'' approaches when only the AGN QLF is considered.  In the current work, we incorporate other observational data, most notably the distribution of AGN in the $(m_{bh},L)$ plane, to move beyond the information in the QLF alone.
\\

The layout of the paper is as follows.  In Section \ref{sec:Ansatze} we first state the three simple Ans\"atze that are used to construct the model.  We then show in Section \ref{sec:OriginQLF} how the quasar $\phi(L)$ can be simply constructed from the galaxy $\phi(m_{*})$ and Eddington ratio distribution $\xi(\lambda)$ via a convolution, show how the parameters of these distribution functions are connected, and make testable predictions of the mass distribution of the host galaxies of quasars selected at different luminosities.  In Section \ref{sec:Data} we review the observed epoch-dependent $\phi_{SF}(m_{*},z)$ and $\phi(L,z)$ functions and then in Section \ref{sec:Comparing} we compare these within the framework of the convolution model to derive interesting conclusions about the duty cycle of AGN.  \\

Up until this point, the model is completely general.  In Section \ref{sec:TWMD}, we then consider different possibilities for the evolution of the $m_{bh}/m_{*}$ ratio and show that one particular choice can explain the observed distribution of luminous quasars in the ($m_{bh},L)$ plane and local scaling relations of black holes in both star-forming and passive galaxies and the possible differences in evolution of active and passive systems.  Section \ref{sec:Dis} presents a discussion of some further implications of both the general model and the specific $m_{bh}/m_{*}$ implementation, differences between $m_{bh}/m_{*}$ and $m_{bh}-\sigma$ the appearance of downsizing within the AGN population, and the pervasive biasses that enter into luminosity-selected samples.  The paper then concludes with a summary section. \\

Throughout the Paper, we will assume a $\Lambda$CDM cosmology, with parameters $\Omega_{M}$=0.3, $\Omega_{\Lambda}$=0.7 and $H_{0}$= 70 km s$^{-1}$ Mpc$^{-1}$. Luminosities will be given in units of erg s$^{-1}$ and will refer to bolometric luminosities, unless specified otherwise. We use the term "dex" to denote the antilogarithm, i.e. $n$ dex = 10$^{n}$.   We also define all distribution functions, i.e. the star-forming galaxy mass function $\phi_{SF}(m_{*})$, the associated star-forming galaxy black hole mass function $\phi_{BH}(m_{bh})$, the AGN luminosity function $\phi(L)$ and the probability distribution of Eddington ratio $\xi(\lambda)$, in log space. This leads to power-law exponents that differ by unity relative to distribution functions defined in linear space. Therefore, the units of $\phi_{SF}(m_{*})$, $\phi_{BH}(m_{bh})$ and $\phi(L)$ are Mpc$^{-3}$ dex$^{-1}$. \\

\section{Ans\"atze} \label{sec:Ansatze}

The essence of this phenomenological approach is to make a limited number of simple Ansätze that allow us to construct a model using analytic techniques, or very elementary numerical modeling, based on straightforward representations of the most important features of the observational data.   These Ans\"atze are in a sense ``assumptions'' and a decisive observational disproof of any of them would largely invalidate the model.  Clearly they are very unlikely to be exactly true. Their value, as the basis for a simple ``toy model'', is that we believe they are likely to be  \textit{broadly} true.\\

The three Ans\"atze in the current work are as follows:

\begin{itemize}
   \setlength{\itemsep}{1pt}
  \setlength{\parskip}{0pt}
  \setlength{\parsep}{0pt}
  \item radiatively efficient AGNs are found in star-forming galaxies,
  \item the probability distribution of the Eddington ratio does not depend on the black hole mass of the system,
  \item the mass of the central black hole is linked to the stellar mass ($m_{bh} \propto m^{\beta}_{*}$), with some scatter, and we will for simplicity set $\beta \sim 1$.

\end{itemize} 

The justification for the first and the third of these has been reviewed in the Introduction and the second is closely related to the assumption of \cite{Air13}.  We will justify these, and the choice of $\beta \sim 1$, further below. Of course, there are a number of other implicit assumptions that are being made: observational data is not wildly wrong, individual black hole and stellar mass measurements are not systematically biased, the cosmological model is more or less correct and so on. We will not consider these further.

\section{The origin of the quasar luminosity function} \label{sec:OriginQLF}

With our Ansätze, we can use our knowledge of the mass function of galaxies to construct the black hole mass function. To do this, we start with the mass function of star-forming galaxies and then impose a black hole to stellar mass ratio $m_{bh}/m_{*}$, with an additional log-normal scatter of $\eta$, to create a black hole mass function in star-forming hosts. This will serve as the basis for the radiatively efficient AGN population. In first part of the paper, we will set the scatter $\eta=0$ for initial analytic simplicity, before reintroducing it with $\eta \sim 0.5$ when we evaluate the model numerically. \\

If radiatively efficient accretion onto central black holes is only occurring in star-forming hosts, the quasar luminosity function can be created from a convolution of the black hole mass function in star-forming galaxies $\phi_{BH}(m_{bh},z) $ with a probability distribution of the Eddington ratio $\lambda$,
\begin{equation} \label{eq:CreateQLF}
\phi (L,z) = \int \phi_{BH} (m_{bh},z) \xi (\lambda,z) d \log\lambda,
\end{equation}
where $\phi(L,z)$ is the resulting QLF, and $\xi (\lambda,z)$ is the probability distribution of the Eddington ratio, $\lambda$, as defined in Equation (\ref{eq:LumWithEd}).

\subsection{Convolution of Schechter functions with power law Eddington ratios}\label{sec:ConvOfSch}

As noted above in Equation (\ref{eq:CreateQLF}), the QLF will, with our Ans{\"a}tze, be a convolution of the black hole mass function $\phi_{BH}(m_{bh})$ (itself derived from the stellar mass function of star-forming galaxies $\phi_{SF}(m_*)$) and the distribution of Eddington ratios $\xi(\lambda)$, which gives the distribution of luminosities for black holes of a given mass.
In this section we look at the general features of the $\xi(\lambda)$ distribution that are needed to produce a double power-law QLF from a Schechter-like mass function and derive the connections that will exist between the parameters of these different functions. \\
\begin{figure*}[htp]
	\centering
  \includegraphics[width=\textwidth]{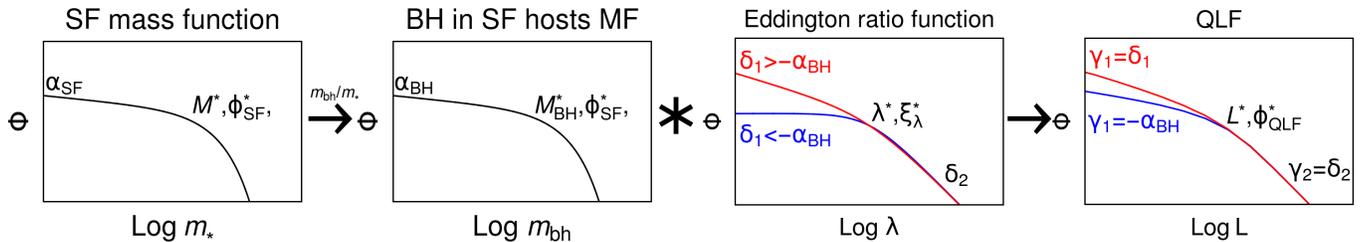}
	\caption{Schematic representation of our procedure to create the QLF. Starting from the star-forming mass function (leftmost) we use an $m_{bh}/m_{*}$ scaling to get the mass function of SMBH in star-forming galaxies, with scatter as required (2$^{nd}$ panel). We convolve it with the Eddington ratio distribution (in the 3$^{rd}$ panel) to create QLF in the rightmost panel. Blue and red Eddington ratio distributions differ in the choice of parametrization of low end slope. The faint end slope of QLF will be same as low mass slope of galaxy mass function or low end slope of Eddington ratio distribution, depending on relative steepness of these two slopes. A short summary of the connections between parameters of the QLF and parameters of the contributing functions is given in Equation (\ref{eq:Master}). }
	\label{fig:CreateLF}
\end{figure*}
First, we immediately note that in a model with a mass-independent $\xi(\lambda)$ and an input Schechter mass function, the only way to produce a power law at the bright end of the QLF (with slope $\gamma_{2}$) is to have a $\xi(\lambda)$ that is also a power law of the same slope at high values of the Eddington ratio.  This power-law will then ensure that there is a high-end power-law in the QLF even as the mass function drops off exponentially. This is shown in the Appendix.  Given the limited dynamic range of the data, other representations of the QLF instead of a double power law are possible, e.g. a log-normal bright end or a modified Schechter function (e.g. \citealp{Hop07}, \citealp{Air12}, \citealp{Vea14}).  
We choose the conventional double power law for analytic simplicity and because it certainly provides a reasonable representation of the available data.   Denoting the slope of the high end of Eddington ratio distribution by $\delta_{2}$ we can thus equate 
\begin{equation} \label{eq:gamma2}
\gamma_{2}=\delta_{2},
\end{equation}
with $\gamma_2$ the bright end slope of the QLF. \\

At the faint end, we may also expect a power-law QLF but now the QLF faint end slope $\gamma_{1}$ will be given by the steeper of the low end slope of $\phi_{BH}(m_{bh})$ and the low end slope of $\xi(\lambda)$, which we denote by $\delta_{1}$ (see also discussion in \cite{Vea14}). Figure \ref{fig:CreateLF} illustrates what is happening with faint end of QLF (with slope) for different Eddington ratio assumptions. 

We can express this as 
\begin{equation} \label{eq:gamma1}
\gamma_{1}= \mbox{max}(-\alpha_{BH},\delta_{1}).
\end{equation}

The natural conclusion of a mass-independent Eddington rate distribution is that the faint end of QLF is set up by either the low end slope of underlying black hole mass function or by the low end slope of the Eddington ratio function. If the logarithmic slope of the $m_{bh}$ vs. $m_{*}$ relation is $\beta$ as above, then the faint end slope of the black hole mass function will be related to that of the star-forming galaxies by $\alpha_{BH} = \alpha_{SF}/\beta$. \\

We note at this point that there is good observational evidence that the faint end QLF slope, $\gamma_{1}$, is similar to the observed low mass slope of the mass function of star-forming galaxies, $\alpha_{SF}$, allowing for the reversal of sign in our definition.
The QLF faint end slope, $\gamma_{1}$ is usually observed to be between 0.3 and 0.9 (\citealp{Has05}; \citealp{Hop07}; \citealp{Air10}; \citealp{Mas12}; \citealp{Ued14}). On the other hand the observed values of $\alpha_{SF}$ range from to -0.6  to -0.2  (\citealp{Bal08}; \citealp{Per08}; \citealp{Pen10}; \citealp{Gon11}; \citealp{Kaj11}; \citealp{Bal12}; \citealp{Lee12}) with most newer studies converging around $\alpha_{SF} \sim -0.4$.\\

The data are therefore consistent with the idea that $\gamma_1 \sim -\alpha_{SF}$. This suggests that $\gamma_1$ is indeed being set by the low end slope of the black hole mass function (and not the low end of the Eddington ration distribution) and further that $\beta \sim 1$.   We will henceforth assume that this is the case, i.e. that $\alpha_{BH}=\alpha_{SF} \approx -\gamma_{1}$. This then means that the low end slope $\delta_1$ of  $\xi(\lambda)$ can take any value that is shallower than this, i.e. $\delta_1 < -\alpha_{SF} \sim 0.4$. 
The high and low end power-laws of the Eddington ratio distribution will meet in a knee at a characteristic Eddington ratio which we denote by $\lambda^*$ at which point the value of $\xi(\lambda)$ has a characteristic value $\xi^*_{\lambda}$.\\

\subsection{The simplest Eddington ratio distribution that reproduces the shape of the QLF}

To further demonstrate our approach we use the simplest function possible for the Eddington ratio distribution that reproduces, analytically, the required broken power law shape of QLF.  We use a "triangular" distribution in which some fraction, $f_{d}$, of objects are active above a certain threshold value  $\lambda^{*}$ with an Eddington ratio distribution that is a single power law with slope $\delta_{2}$, while (1-$f_{d}$) of objects are completely inactive.  Effectively this sets the low end slope $\delta_{1} \sim - \infty$.\\

The exact functional form of $\xi(\lambda)$ can then be written as 
\begin{equation} \label{Ed}
\xi(\lambda)=\frac{dN}{N d \log \lambda}=  \xi^{*}_{\lambda} \left( \frac{\lambda}{\lambda^{*}}\right) ^{-\delta_{2}}, \hspace{1 cm}\lambda > \lambda^{*}, 
\end{equation}
where $\xi^{*}_{\lambda}$ in constrained by requirement that all of the objects have to be either active or inactive, so $\xi$ in this case has to integrate to a duty cycle $f_{d}$.\\

In an Appendix we show that the knee of the QLF, $L^{*}$, will be related to the parameters of the original distribution functions through a formula describing the \textit{population} which is analogous to Equation (\ref{eq:LumWithEd}) of \textit{individual} black holes, i.e.
\begin{equation} \label{eq:Lstar}
L^{*}=10^{38.1} \cdot \lambda^{*} \cdot M^{*}_{BH} \cdot \Delta_L(\gamma_2),
\end{equation}
where the $\Delta_L$ factor denotes a small multiplicative factor that is weakly dependent on $\gamma_{2}$ and varying by less than 0.15 dex. \\

In the same Appendix, we also show that the $\phi^{*}_{QLF}$ normalization of the QLF, i.e. the value of  $\phi_{L} $ at $L = L^*$ will be linked to the normalization of the galaxy mass function $\phi^{*}_{SF}$ and the normalization of the Eddington ratio distribution, $\xi^{*}_{\lambda}$, 

\begin{equation} \label{eq:phi}
\phi^{*}_{QLF} = \phi^{*}_{SF} \cdot \xi^{*}_{\lambda}\cdot \Delta_\phi(\gamma_2).
\end{equation}
where the $\Delta_\phi$ denotes once again a small multiplicative factor, weakly dependent on $\gamma_{2}$ and varying by less then 0.15 dex. This is not surprising and is stating the fact that the normalization of objects at given luminosity $L^{*}$  is connected with the normalization of the contributing functions at $M^{*}_{BH}$ and $\lambda^{*}$. Simply put, if there are more objects at mass $M^{*}$ and more of them are active at $\lambda
^{*}$, then we expect also more objects with corresponding luminosity $L^{*}$. \\

\subsection{Broken power law Eddington ratio distribution and generalized duty cycle}\label{sec:33}

Even though the triangular distribution of $\xi(\lambda)$ is a simple and analytically tractable one, it is unlikely to describe the real AGN population. In the remainder of this paper we will adopt a more realistic broken power law distribution of Eddington ratio given by 
 
\begin{equation} \label{Ed2}
\xi(\lambda)=\frac{dN}{N d \log \lambda}=  \frac{\xi^{*}_{\lambda}}{(\lambda/\lambda^{*})^{\delta_{1}}  +(\lambda/\lambda^{*})^{\delta_{2}} }, 
\end{equation}
where we set $\delta_{1}$=0 for further analysis. This function fits both observations and hydrodynamical simulations better, which show no sudden cutoff in the Eddington ratio distribution at some single value (e.g. \citealp{Nov11}, \citealp{Kel13}). This distribution diverges logarithmically at the low $\lambda$ end.  Since the integral of $\xi(\lambda)$ will therefore reach unity at some low value of $\lambda << \lambda^*$, all black holes are ``active'' at some very low level and we need no longer consider ``inactive'' ones. \\

As discussed in Section \ref{sec:ConvOfSch}, this distribution will also reproduce the same shape of the QLF as the "triangular" distribution just discussed, provided that $\delta_{1}<-\alpha_{BH}$. \\

 We note that such a distribution, (with sharp break at $\lambda^{*}$, instead of "smooth" version above), would naturally arise if individual AGN are boosted to some initial Eddington ratio above $\lambda^*$ (the distribution of which is given by $\gamma_2$) and thereafter decay exponentially with a constant decay time $\tau$. In this case, the concept of "duty cycle"  $f_d$ in the previous section is replaced by the value of $\xi^*_{\lambda}$.   If the ``boost plus decay'' scenario is relevant, then it is easy to see that, if AGN are activated at some fractional rate $\eta$ per star-forming galaxy per unit time (e.g. \citealp{Hop05}), then 

\begin{equation} \label{fd}
\xi^*_{\lambda} = \eta \cdot \tau,  
\end{equation}
so that $\xi^{*}_{\lambda}$ is the fraction of black holes that are activated in the time interval corresponding to their subsequent decay time. This is quite a useful definition of duty cycle.  \\

Of course, other scenarios could also produce this $\xi(\lambda)$ distribution and a more general ``duty cycle'' could be defined as the fraction of black holes accreting above $\lambda^*$ (as in the triangular distribution above). In this particular case, as shown in the Appendix, this definition of duty cycle would be given by  

\begin{equation} \label{eq:ap31}
f_{d}= \frac{ \xi^{*}_{\lambda}  }{\delta_{2}\ln(10)}    
\end{equation}

The point is that the value of $\xi^*_{\lambda}$ is a good indicator of a generalized duty cycle.\\

If $\delta_{1}< - \alpha_{BH}$, then the exact choice $\xi(\lambda)$ below $\lambda^*$ is of no great importance to the convolution model and the connections between the parameters, which we can derive analytically in the simplified model as in Equations (\ref{eq:gamma2}), (\ref{eq:gamma1}), (\ref{eq:Lstar}) and  (\ref{eq:phi}) will still hold.  Specifically, we can write

\begin{equation} \begin{split} \label{eq:Master}
\gamma_{2} &= \delta_{2}, \\
\gamma_{1}&= \mbox{max}(-\alpha_{BH},\delta_{1}),\\
L^{*} &=10^{38.1} \cdot \lambda^{*} \cdot M^{*}_{BH} \cdot \Delta_L(\delta_1,\gamma_2),\\
\phi^{*}_{QLF} &= \phi^{*}_{SF} \cdot \xi^{*}_{\lambda}\cdot \Delta_\phi(\delta_{1},\gamma_2),
\end{split}\end{equation}
where we have now also explicitly shown a $\delta_{1}$ dependence in the $\Delta$ factors, because this corrective factor will also depend on our exact choice of low slope of the Eddington ratio distribution. Now using the $m_{bh}/m_{*}$ relation, we can also connect the observed $L^{*}$ of the QLF to the $M^{*}$ of the star-forming galaxy mass function with 
\begin{equation} \label{eq:MasterL}
L^{*}=10^{38.1} \cdot \lambda^{*} \cdot M^{*} \cdot \left( \frac{m_{bh}}{m_{*}} \right) \cdot \Delta_L(\delta_1,\gamma_2).
\end{equation}

\subsection{Predictions of the convolution model}\label{sec:Pred}

\begin{figure*} 
	\centering
  \includegraphics[width=0.99\textwidth]{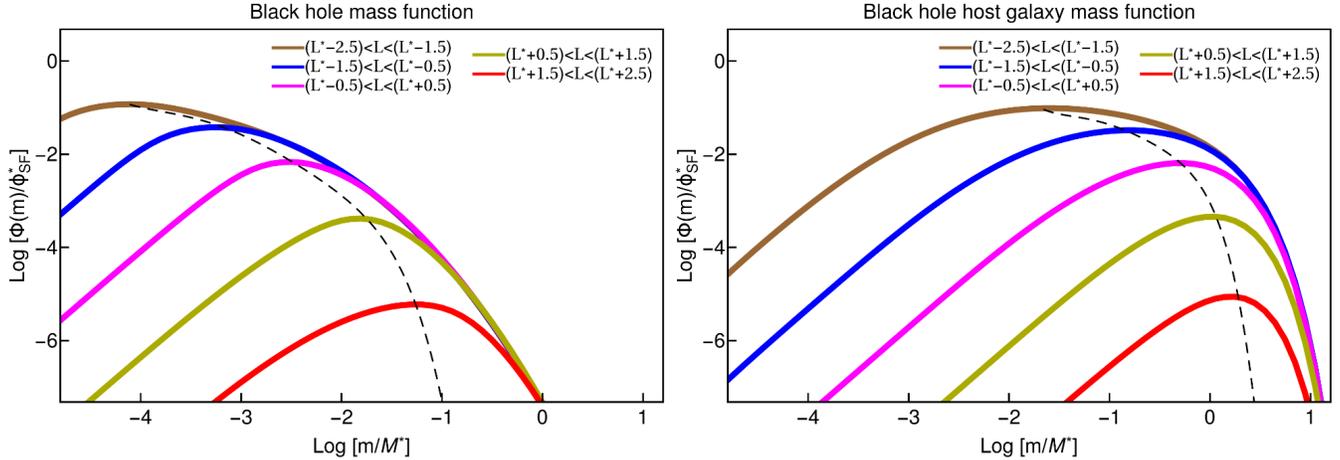}
	\caption{Expected shape of the black hole and the black hole host galaxy mass function, selected in different AGN luminosity bins. Mass are ploted relative to the Schechter $M^{*}$ of the galaxy population ($M^{*}\sim 10^{10.8} M_{\Sun}$) and $\Phi$ relative to the $\phi^{*}_{SF}$ of the galaxy population. All luminosities are relative to the $L^{*}$ of the AGN luminosity function.  The black dashed line is connecting the masses where the mass functions reach maximum value.   } 
\label{fig:MFAGN}
\end{figure*}

If the QLF is indeed produced by the convolution described in the previous two sections, then we can immediately see in general terms how the mass distribution of AGN, or host galaxies of AGN, selected at a given luminosity will vary with that selection-luminosity. This is shown in Figure \ref{fig:MFAGN} where we show the mass functions of galaxies and black holes.  The masses and number densities are normalized respectively by the $M^{*}$ and $\phi^{*}_{SF}$ of the input Schechter mass function of the star-forming galaxies. These distributions are plotted for different AGN luminosities relative to the knee luminosity $L^*$ in the quasar luminosity functions (see also Equation  (\ref{eq:QLFF})).   To generate this plot we used the model for $\xi(\lambda)$ considered in the previous section, i.e. with a flat $\xi$ below $\lambda^*$ and with 0.5 dex scatter in the black hole to stellar mass relation. With this normalization the figures applies at all redshifts. Several points should be noted on this Figure. \\

First, all of the mass functions (of both black holes and host galaxies) always show a peak at some mass.  This is quite unlike the input mass function of star-forming galaxies, and thus also of black holes, which increase monotonically to lower masses.   The peak arises because the low mass end of the input mass function is modulated by the high $\lambda$ part of $\xi(\lambda)$, simply because it is the highest Eddington ratios that pull the lowest mass black holes (and hosts) into a given luminosity bin.  The low mass end of these mass distributions should therefore have a slope of $\gamma_2 + \alpha_{SF}$, i.e. with $\alpha_{SF} \sim -0.4$ and $\gamma_2 \sim 2$ (see below), we predict a low mass slope of +1.6.  This is a quite generic and robust prediction of the convolution model and is independent of the choice of $\delta_{1}$, the low $\lambda$ behavior of $\xi(\lambda)$ discussed above.  This is because for most luminosities, the \textit{high} mass end of the mass functions in Figure \ref{fig:MFAGN} is dominated by the exponential fall-off of the input mass function. \\

Second, at high luminosities, above $L^*$, the effect of the AGN luminosity on the host galaxy mass function is to change the numbers of hosts, but not to change the peak mass or, to first order, the shape of the mass function.  This is because of the steep exponential fall off in the input mass function of galaxies above $M^{*}$.  Above $L^*$, the peak host galaxy mass is always close to $M^*$ because there are so few more massive galaxies. The numbers of $M^{*}$ galaxies at the peak is determined by how high Eddington ratios are required to bring these objects into the luminosity range in question.  For these high AGN luminosities, the shape of the mass function of (star-forming) hosts is very similar to that of passive galaxies except that the faint end slope is significantly steeper ($\alpha_{SF} + \gamma_2$, with $\gamma_2 \sim 2$) that of the passive galaxies which is given by $\alpha_{SF} + 1$  (see \cite{Pen10}). In mathematical form, the mass function of galaxies at any AGN luminosity above $L^{*}$ will have a Schechter shape
\begin{equation}
\phi(m_{*}; L > L^{*}) \propto \left( \frac{m_{*}}{M^{*}} \right)^{\alpha_{SF}+\gamma_{2}} \exp \left( -\frac{m_{*}}{M^{*}}  \right)   .
\end{equation}

At lower luminosities, well below $L^*$, the behaviour changes and a region of intermediate slope appears. For black hole masses between $(10^{38.1})^{-1} (m_{bh}/m_{*})^{-1} (\lambda^*)^{-1} L$ and $(10^{38.1})^{-1} (m_{bh}/m_{*})^{-1} (\lambda^*)^{-1} L^{*}$, the mass function will have the slope given by $\alpha + \delta_{1}$. At very low masses we will again always recover the slope of $\alpha_{SF} + \gamma_{2}$.\\

The location of the peak in the black hole mass function therefore depends on the slopes of both underlying distributions (Eddington ratio and galaxy mass function), i.e. on the sign of $\alpha + \delta_{1}$.  Not surprisingly, this intermediate zone also appears in the host galaxy mass function with the same slope. This produces a peak host galaxy mass of

\begin{equation}\begin{split}
m_{peak} &\sim \frac{M^{*}}{10^{38.1 }m_{bh}\lambda^{*}} L  \hspace{1.1 cm}  |\alpha_{SF}|>|\delta_{1}| \\
m_{peak} &\sim \frac{M^{*}}{10^{38.1}  m_{bh}\lambda^{*}} L^{*}  \hspace{1 cm} |\alpha_{SF}|<|\delta_{1}|.
\end{split}\end{equation}
This difference in behaviour can be understood as follows: In the $|\alpha_{SF}|>|\delta_{1}|$ case the dominant contribution to the luminosity function will come from low mass objects accretion at high Eddington ratio which are more numerous than high mass objects accreting at low Eddington ratios. In the second case, roles are simply reversed and main contributor to the QLF will be high mass AGNs with low Eddington ratios. \\

If black hole masses were very tightly correlated with host galaxy masses, then clearly the same behavior would be seen in the black hole mass function(s) since these would always be a simple (mass-)scaling of the galaxy mass function(s).  However, the effect of scatter in the black hole to host galaxy mass relation is quite marked. This is visible in Figure \ref{fig:MFAGN}.  The reason is clear: the mass function of black holes will now be much smoother than that of the host galaxies since the log normal scatter smooths out the sharp exponential drop of the galaxies at high masses (recall that Figure 2 used a 0.5 dex scatter).  Of course, it might be argued that the black hole mass is somehow more fundamental, and that the galaxy mass function should be obtained by smoothing it.  However, we know the galaxy mass function quite well, and it has an exponential cutoff (see \citealp{Pen10}, \citealp{Bal12}).  We do not, at this stage, know the underlying mass function of black holes in star-forming galaxies. With this scatter, there is a much smoother variation of the peak mass with AGN luminosity. \\

The dashed lines show the variation of the peak mass with luminosity.  The fact that the difference between these dashed curves on the two panels decreases with luminosity illustrates the well known ``Lauer effect'' \citep{Lau07} whereby the typical objects at high AGN luminosities will generally have been scattered above the mean black hole host mass relation.  It should be noted, however, that there is no change in the low mass slope of the mass function(s) due to scatter and this remains a very robust prediction of our model. \\

It is worth noting in Figure \ref{fig:MFAGN} that the peak of the black hole mass distribution varies quite weakly with luminosity, i.e. a change in luminosity of 1 dex is associated with a smaller increase in the peak $m_{bh}$.  This means that we will generally see higher Eddington ratios in higher luminosity quasars even though the Eddington ratio distribution is taken to be strictly independent of black hole mass.  We return to this point in Section \ref{sec:plane} below. \\

We stress that the (solid) curves in Figure \ref{fig:MFAGN} that show the mass functions of AGN host galaxies (with masses normalized to the Schechter $M^*$) at different AGN luminosities (computed relative to the knee of the luminosity function) are an easily testable prediction of the convolution approach, modulo the effects linked to the choice of $\alpha$ and $\delta_1$ discussed above, which can shift the peak of $\phi(m)$.

\section{Data}\label{sec:Data}

We next turn to demonstrate how our model relates to the observations of the mass function of star-forming galaxies and QLF. In this section, we will estimate the redshift evolution of parameters that describe these two populations.  

\subsection{Mass function of star-forming galaxies} \label{subsec:MSF}

As explained previously, we need to know the mass function of star-forming galaxies to deduce the SMBH mass function in star-forming galaxies, $\phi_{BH}(m_{bh})$, which is an integral part of predicting the QLF in Equation (\ref{eq:CreateQLF}). We also want to verify the predictions of \cite{Pen10} and \cite{Pen12} for the time evolution of the parameters of the galaxy mass function (e.g. equation B3 from \citealp{Pen12}). \\ %

\begin{figure}[htp]
	\centering
  \includegraphics[width=0.49\textwidth]{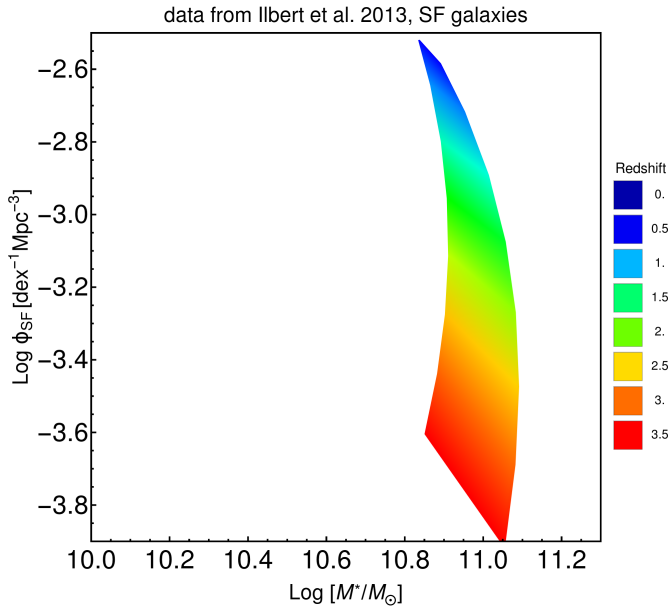}
	\caption{Evolution of $\phi^{*}_{SF}$ and $M^{*}$ parameters in the Schecter mass function of star-forming galaxies. Data shows minimal evolution of $M^{*}$ until at least $z \approx$ 2.5, but significant evolution in $\phi^{*}_{SF}$.} 
	\label{fig:ModLum}
\end{figure}

In order to do this, we fit the data for star-forming galaxies from \cite{Ilb13} at all redshifts with a single Schechter function in which the parameters ($\phi^{*}_{SF}\equiv \phi^{*}_{SF}(z) $, $M^{*}\equiv M^{*}(z)$) are smoothly varying functions of redshift. Fitting all the data in this fashion, instead of fitting at single redshifts, enables us to follow better the evolution of the parameters. It also reduces the sensitivity to small variations which may compromise fits at a single redshift and add to the degeneracies between parameters. This procedure of simultaneously fitting all of the data of the galaxy mass function at different redshifts is analogous to the standard procedure often used in determining the evolving QLF (e.g. \citealp{Hop07}, \citealp{Ued14}). The functional form that we use for the redshift evolution of the galaxy mass function is
\begin{equation}\begin{split} \label{eq:SF}
\phi_{SF}(m_{*},z) &= \frac{dN}{d \log m_{*}} \\
&=\phi^{*}_{SF}(z) \left( \frac{m_{*}}{M^{*}(z)}\right)^{\alpha_{SF}} \exp\left(- \frac{m_{*}}{M^{*}(z)} \right), 
\end{split}\end{equation}
where we model the redshift dependence as 
\begin{equation}\begin{split} \label{eq:redMphiSF}
\log \phi^{*}_{SF}(z)& = a_{0}+a_{1}\kappa+a_{2}\kappa^{2}+a_{3}\kappa^{3},\\
\log M^{*}(z) &= a_{0}+b_{1}\kappa+b_{2}\kappa^{2} +b_{3}\kappa^{3},
\end{split}\end{equation}
with $\kappa \equiv \log (1+z)$.
The slope of the low mass end, $\alpha_{SF}$, was kept constant at the local value of $\alpha_{SF} = -0.4$ as there is considerable evidence that there is little or no change in the low mass slope for the redshift range considered \citep{Pen14}.\\

\begin{deluxetable}{lc}
\tablecaption{Evolution of star-forming galaxy mass function, data from \cite{Ilb13}}
\tablewidth{0pt}
\tablehead{parameter & fixed $\alpha_{SF}$  = -0.4  }
 $a_{0}$	& -2.55 $\pm$ 0.04	  \\
 $a_{1}$	& -0.26 $\pm$ 0.06 	 \\
 $a_{2}$& -1.6 $\pm$ 0.12	  \\
 $a_{3}$	& -0.88 $\pm$ 0.16 \\
 $b_{0}$&	10.9 $\pm$ 0.04 \\
 $b_{1}$	& -0.53$\pm$ 0.11 \\
 $b_{2}$	& 3.36 $\pm$ 0.11  \\
 $b_{3}$	&  -3.75 $\pm$ 0.13	 \\
 \label{tab:qlf}
\end{deluxetable}

The evolution of parameters of mass function in Figure \ref{fig:ModLum} confirms that $M^{*}$ is more or less constant up until at least redshift 2.5, i.e. it verifies the Ansatz of \cite{Pen10} which establishes a single quenching mass scale $M^{*}$ that does not change with redshift.  It is important to stress that our results will not depend critically on the exact functional evolution of $M^{*}$ above $z\gtrsim 2.5$, but will use the by now well established observational fact that $M^{*}$ does not change at $z \lesssim 2.5$ and that the evolution in the galaxy population since that time is associated with a ``vertical'' evolution in $\phi^*_{SF}$. We have also performed this analysis with the dataset from \cite{Muz13} and the compilation of galaxy mass functions from \cite{Ber13} and our conclusions presented in the rest of this paper do not change. \\

\subsection{Quasar luminosity function}

\cite{Hop07} combined measurements of quasar luminosity functions in different bands, fields and redshifts in order to characterise the bolometric QLF at epochs $0 < z < 6$. In their work, the best fit luminosity-dependant bolometric correction and luminosity and redshift-dependent column density distributions are used in order to construct an estimate of the bolometric QLF which should be consistent with all of the various individual surveys.   Although now several years old, we believe that this synthesized QLF compilation remains the most comprehensive and is used as the basis of the current paper.\\

We proceed with fitting their tabulated data with a double power law QLF, as given by Equation (\ref{eq:QLFF}).  We fit both at each redshift individually, and carry out a ``full fit'' where the parameters (i.e. $\phi^{*}_{QLF}$, $L^{*}$, $\gamma_{1}$ and $\gamma_{2}$) are all constrained to be smoothly varying functions of redshift, i.e. adopting a similar approach as we used earlier in our fitting of the galaxy mass function in Section \ref{subsec:MSF}.  This ``full'' fitting of a redshift-dependent double power-law is essentially the same as the approach used in \cite{Hop07}, with one important difference.  In order to avoid degeneracies in the fits it is necessary, both here and in \cite{Hop07}, to fix one or more of the parameters.  \cite{Hop07} chose to require a redshift-independent $\phi^{*}_{QLF}$ at a constant value.  We now know that $\phi^{*}_{SF}$ of the galaxy population changes significantly over the redshift range of interest and, as developed in the previous section, the relationship of $\phi^{*}_{QLF}$ relative to $\phi^{*}_{SF}$ is of great interest in the context of the duty cycle.  In contrast, $\gamma_{1}$ is set, in our convolution model, by the low mass slope $\alpha_{SF}$ of the mass function of star-forming galaxies. As mentioned before there is not much evidence (see \citealp{Pen14}) that this changes significantly, if at all, over the redshift range of interest, nor compelling evidence for a change in $\gamma_{1}$.  Therefore, we choose to have a redshift-independent faint end slope of the luminosity function, $\gamma_{1}$ and to allow $\phi^{*}_{QLF}$ to vary with redshift.   It should be noted that our fitting procedure is the same as a ``luminosity and density evolution'' model (e.g. \citealp{Air10}) except that we are allowing the bright end slope $\gamma_2$ to vary.  We do this because, as we have seen, $\gamma_2$ is set by the high $\lambda$ behavior of the Eddington ratio distribution $\xi(\lambda)$.\\

\begin{figure*}[htp]
	\centering
  \includegraphics[width=\textwidth]{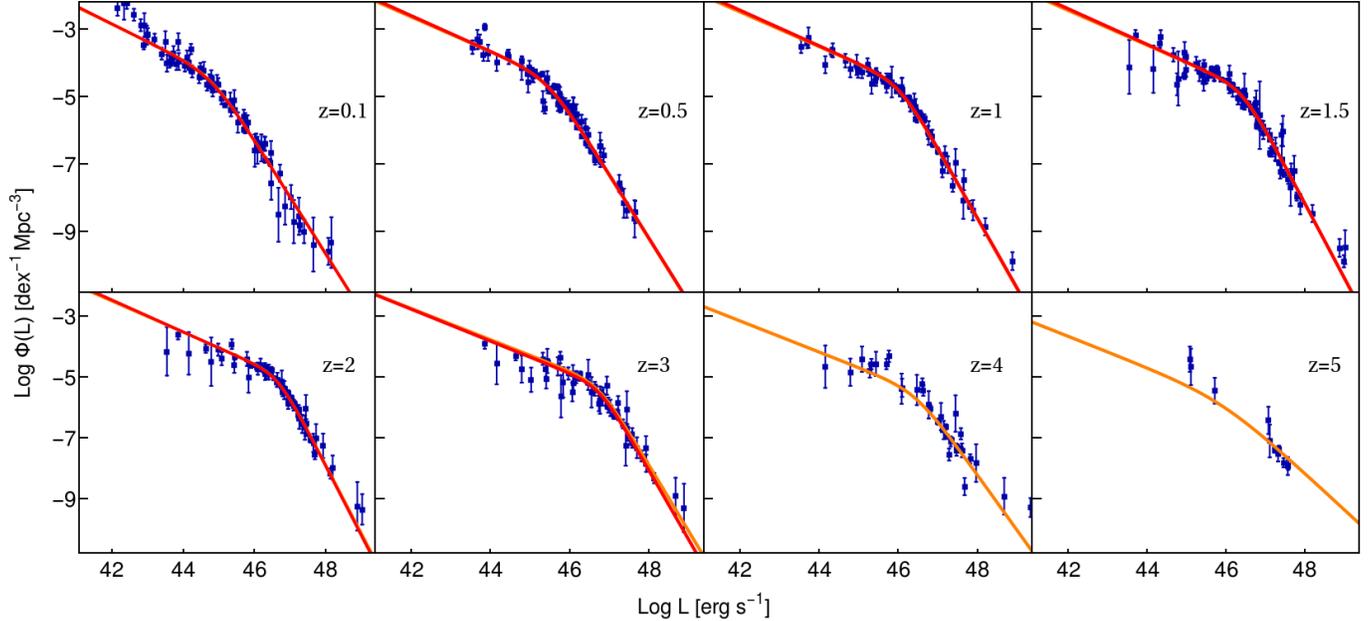}
	\caption{Fits to the bolometric dataset compiled in \cite{Hop07}, with the broken power law fit of Equation (\ref{eq:QLFF}). The orange curve shows our ''full fit" done with parametrization in Equation (\ref{eq:RedEv1}). Red curve shows fit in which redshift dependence of $\phi^{*}_{QLF}$ is set to be exactly the redshift dependence of  $\phi^{*}_{SF}$.} 
	\label{fig:QLF}
\end{figure*}

The parameters in the luminosity function are allowed to vary as 
\begin{equation}\begin{split} \label{eq:RedEv1}
\mbox{log } \phi^{*}_{QLF}& = c_{0}+c_{1}\kappa+c_{2}\kappa^2+c_{3}\kappa^3,\\
\mbox{log } L^{*} &= d_{0}+d_{1}\kappa+d_{2}\kappa^{2}+d_{3}\kappa^{3}+d_{4}\kappa^{4} ,\\
\gamma_{1} &= e_{0},\\
\gamma_{2} &= f_{0}+f_{1}z+f_{2}z^{2},
\end{split}\end{equation}
with again $\kappa = log (1+z)$.  We follow \cite{Hop07} in fitting $\gamma_1$ and $\gamma_2$ as a polynomial in $z$ (rather then $\kappa = log (1+z)$ ), but this choice is not of great importance.\\

\begin{deluxetable}{lcc}
\tablecaption{Best fits to QLF data from \cite{Hop07}}
\tablewidth{0pt}
\tablehead{parameter &``full fit" & with $\phi^{*}_{QLF} \propto \phi^{*}_{SF}$  }
 $c_{0}$	& -4.35 $\pm$ 0.06	& -1.79 $\pm$ 0.04 \\
 $c_{1}$	& 0.059 $\pm$ 0.07 	& - \\
 $c_{2}$& -3.2 $\pm$ 0.3	& - \\
 $c_{3}$	& 1.73 $\pm$ 0.2 & -\\
 $d_{0}$&	44.67 $\pm$ 0.06 &  44.67 $\pm$ 0.04\\
 $d_{1}$	& 4.02$\pm$ 0.07	&  4.09 $\pm$ 0.09 \\
 $d_{2}$	& 3.78 $\pm$ 0.08&  3.5 $\pm$0.07  \\
 $d_{3}$	&  -4.68 $\pm$ 0.1	& -3.85 $\pm$ 0.07  \\
 $d_{4}$& -5.7  $\pm$ 0.2 & -5.98 $\pm$ 0.15 \\
 $e_{0}$& 0.5 $\pm$ 0.03	& 0.52 $\pm$ 0.04  \\
$f_{0}$ & 1.64  $\pm$ 0.05	&  1.63 $\pm$ 0.06  \\
 $f_{1}$	&	0.54  $\pm$ 0.05	& 0.56 $\pm$ 0.06 \\
 $f_{2}$ &  -0.12 $\pm$ 0.01	&  -0.106 $\pm$ 0.03 
 \label{tab:qlf}
\end{deluxetable}

Additional degrees of freedom were added to the fit until we can find no appreciable quantitative difference in the quality of the fit. The values of the double power-law parameters in the smoothly varying ``full-fit'' are given in Table \ref{tab:qlf} and the fits are shown with orange curves in Figure \ref{fig:QLF}.  \\

\begin{figure*}
	\centering
  \includegraphics[width=0.99\textwidth]{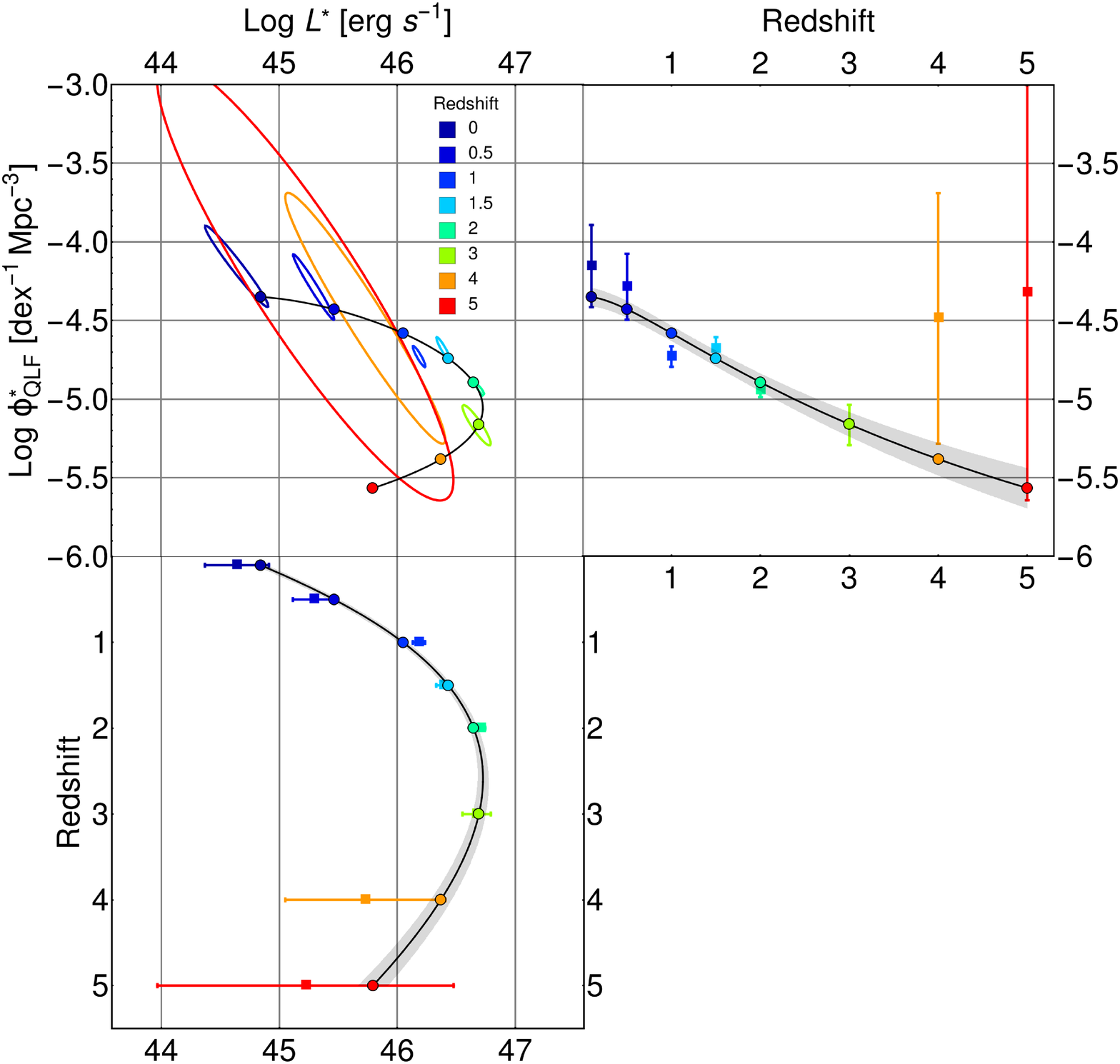}
	\caption{\textit{Top left:} Redshift evolution of QLF parameters $L^{*}$ and  $\phi^{*}_{QLF}$ in our "full fit". The contours are showing 1-$\sigma$ allowed regions of parameter space for fits which were made  with data at each individual redshift. The filled circles are the result of a global fit, for which the resulting QLF is shown in Figure \ref{fig:QLF}. Uncertainty contours for this fit are not shown here for clarity. \textit{Top right panel:} Projection showing explicitly the change of normalization $\phi^{*}_{QLF}$ with redshift.  \textit{Bottom left:} Projection showing explicitly change of $L^{*}$ with redshift. } 
	\label{fig:ParEvoFull}
\end{figure*}

The redshift dependences of $\phi^{*}_{QLF}$ and $L^{*}$ are shown in the panels of Figure \ref{fig:ParEvoFull} and we discuss these results in the next Section. 

\begin{figure*}
	\centering
  \includegraphics[width=0.99\textwidth]{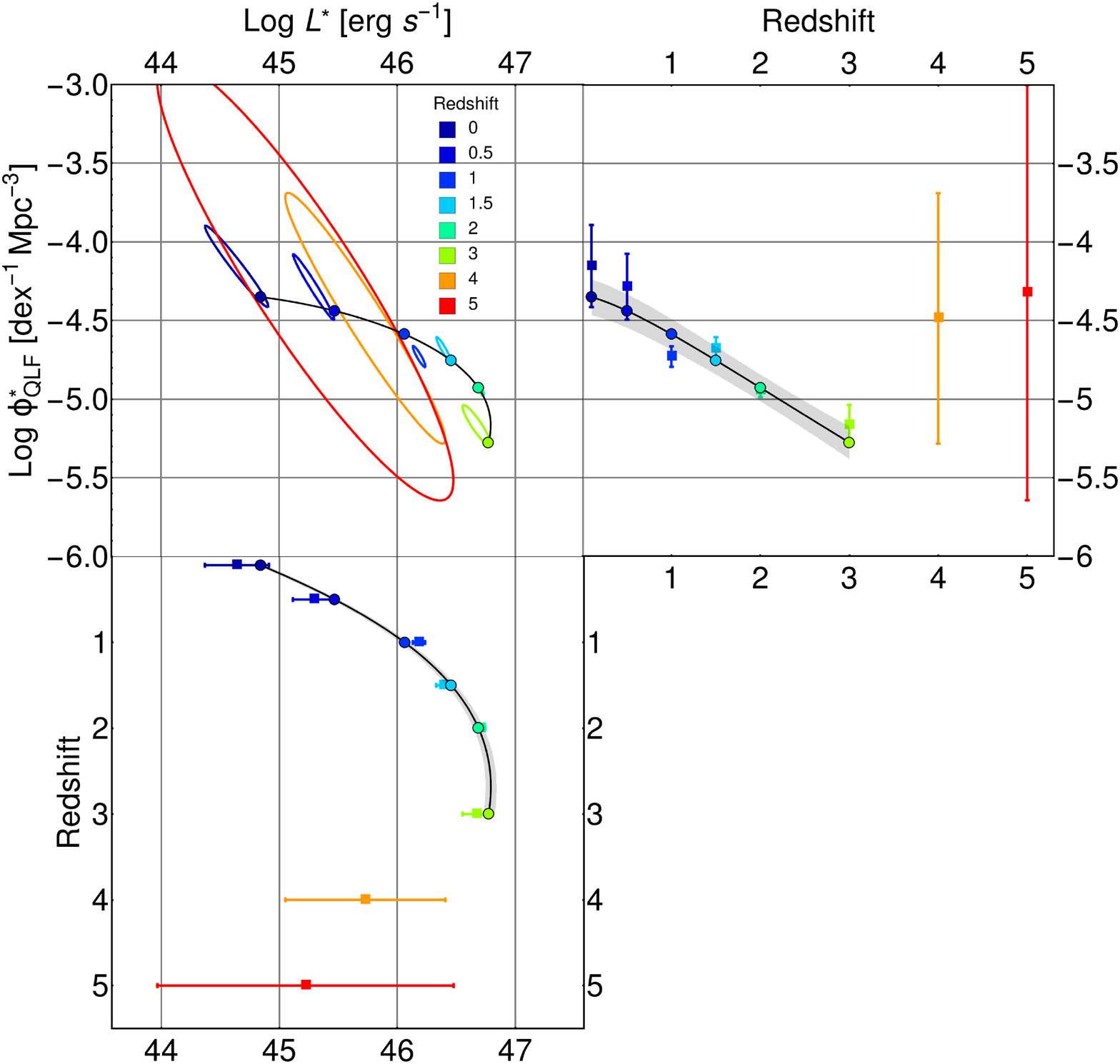}
	\caption{Same as Figure \ref{fig:ParEvoFull} for QLF fit at redshifts $z \leq 3$, in which we demanded that $\phi^{*}_{QLF} \propto \phi^{*}_{SF} $, i.e. that redshift dependence of theses two variables is the same.} 
	\label{fig:ParEvoPeng}
\end{figure*}

\begin{figure}
	\centering
  \includegraphics[width=0.49\textwidth]{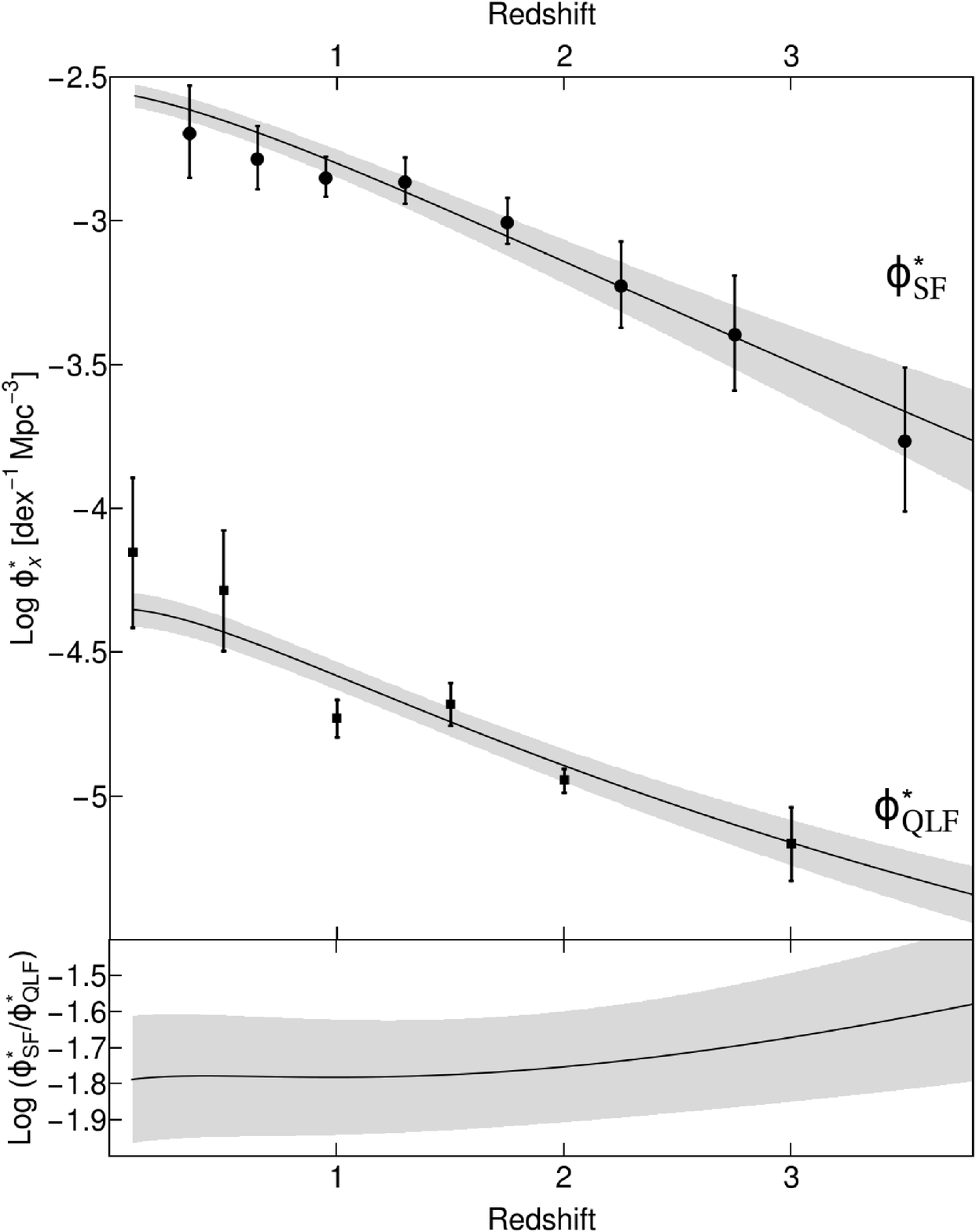}
	\caption{\textit{Top:} Comparison of the redshift evolution of the normalization of star-forming galaxy mass function and normalization of quasar luminosity function. Black circles show the values of $\phi^{*}_{SF}$ which were determined by fitting the data from \cite{Ilb13}  at a single redshift. Black squared show the values of $\phi^{*}_{QLF}$ which were determined by fitting the QLF data at a single redshift and are also shown in Figures \ref{fig:ParEvoFull} and \ref{fig:ParEvoPeng}. \textit{Bottom:} Redshift evolution of the $\phi^{*}_{SF}/\phi^{*}_{QLF}$. This ratio is remarkably constant over redshift range for which data is available. } 
	\label{fig:ParEvoPeng2}
\end{figure}

\section{Results from comparing the evolution of the QLF and the galaxy mass function }\label{sec:Comparing}

\subsection{The evolution of $\phi^*_{QLF}$}

The first result is that we notice a strong similarity between the observed redshift dependence of $\phi^*_{QLF}$ in Figure \ref{fig:ParEvoFull} and the observed $\phi^*_{SF}$ in Figure \ref{fig:ModLum}.  Both drop by $\sim$ 0.5 dex by redshift 2.   Their relative evolution is explicitly compared in the bottom right panel of Figure \ref{fig:ParEvoPeng2}, which shows that a constant ratio between $\phi^*_{QLF}$ and $\phi^*_{SF}$ is perfectly consistent with the data.  We note that the fact that the density normalization of the QLF decreases slowly with redshift has been seen in several previous analyses, for instance in the "luminosity and density evolution" model of \cite{Air10}, which has a 0.4 dex decrease in normalization between redshifts 0 and 2, \cite{Cro09} shows a very similar decrease of normalization in his "luminosity and density evolution" model, while \cite{Has05} show a decrease in normalization of 0.5 dex between redshift 0.6 and 2.4. Here we highlight the striking similarity of this behavior to the observed decline in $\phi^*_{SF}$.\\ 

Referring back to Section \ref{sec:33}, a constant ratio between $\phi^*_{QLF}$ and $\phi^*_{SF}$ implies a constant duty cycle of the black holes in the context of our convolution model.
To explore this further, we now repeat the fitting procedure but set the evolution of $\phi^*_{QLF}$ to be exactly that of $\phi^*_{SF}$, i.e. we impose that the evolution of the QLF normalization is the same as the evolution of the normalization of star-forming galaxies in the redshift range where we have $\phi^*_{SF}$ available (z$<$3.5), while the constant multiplicative offset between these parameters remains a free parameter to be determined by the fitting procedure, i.e. 

\begin{equation}\label{eq:RedEv2}
 \phi^{*}_{QLF} \propto \phi^{*}_{SF}.
\end{equation}

The results of this fitting are given in Table \ref{tab:qlf} and
the resulting QLF is shown with red curve in Figure \ref{fig:QLF}. It can be seen that the fit is extremely good, being only marginally worse then the full fit of \cite{Hop07} ( $\chi^{2}_{this\mbox{ } work}/d.o.f. = 2.1$; $\chi^{2}_{Hopkins}/d.o.f. = 2.0$), both of which have the comparable number of free parameters. The main driver for the slightly worse fits in our work is the deviation of the fit from data for low luminosities at low redshifts.  The data in this regime may be contaminated by contributions from the stellar populations of the hosts, as discussed in \cite{Hop07}. \\

The parameter evolution in this "$\phi^{*}_{SF}$-matched" fitting are shown as linked filled circles in Figure \ref{fig:ParEvoPeng}.  As one can see, the points are typically situated on the edges of the individual 1-$\sigma$ contours, which is to be expected given that  $\chi^{2}/d.o.f. = 2.1$. We conclude that the change of normalization of the QLF is perfectly consistent with the change of normalization of the star-forming galaxy mass function over the entire redshift range for which we have the measurements of normalization of the star-forming galaxy mass function. \\
 
The fact that as far as we can tell the $\phi^{*}_{SF}(z)$ and $\phi^{*}_{QLF}(z)$ track each other throughout cosmic time (at least since $z \sim 3$, and quite possibly since earlier epochs also, is an interesting result. It means that the factor which is connecting these two quantities, which in the convolution model is $\xi^*_{\lambda}$ (slightly modified by $\Delta_{\phi}$), remains constant. \\

As we have discussed above, this suggests that the general \textit{ "duty cycle", $f_{d}$}, stays constant over cosmic time (e.g. Equations (\ref{eq:phi}) and  (\ref{eq:ap31}))(see also \citealp{Con13}).  Clearly, this would not be the case for other definitions of ``duty cycle'' that are based, for instance, on the fraction of black holes radiating above some luminosity or accreting above some  Eddington ratio, if $\lambda^{*}$ or $L^*$ evolves with time, as it does (see below), but we believe that our definition of duty cycle is the most natural one (as discussed above).\\  

\subsection{The evolution of $L^*_{QLF}$}

The other striking feature of the quasar luminosity function is the strong redshift evolution of $L^{*}$, which increases by almost two orders of magnitude back to $z \sim 2$.  At higher redshifts, the $L^{*}(z)$ certainly flattens out and probably declines, although this cannot be established at high significance. The strong initial rise with redshift is seen independent of whether we use the full fit, or the fit constrained to have $\phi^{*}_{SF}(z)$ and $\phi^{*}_{QLF}(z)$ tracking each other.\\

In our convolution model, the steep rise in $L^{*}(z)$ (see Equation (\ref{eq:MasterL}))
could have been caused in principle by one or more of (a) an evolution of $\lambda^{*}$, (b) an evolution of $M^{*}$ of the galaxy mass function or (c) an evolution of the mass ratio $m_{bh}/m_{*}$.\\

As discussed in Section \ref{subsec:MSF}, we know that the characteristic $M^{*}$ of the galaxy population does not change, especially in the redshift range where increase of $L^{*}$ is most prominent ($z \leq 2$), so case (b) will not apply.  There is however complete degeneracy between cases (a) and (c), i.e. the distribution of specific accretion rates of AGN and the black hole to stellar mass ratio.  This is clear in Equation  (\ref{eq:MasterL}) and as has been pointed out by e.g. \cite  {Vea14}.  This degeneracy can only be broken if we have information on the black hole masses of the AGN.  We will explore this in the next section of the paper.\\

\section{Testing the model with black hole mass data}\label{sec:TWMD}

\subsection{The quasar mass-luminosity plane} \label{sec:plane}

\begin{figure*}
	\centering
  \includegraphics[width=0.95\textwidth]{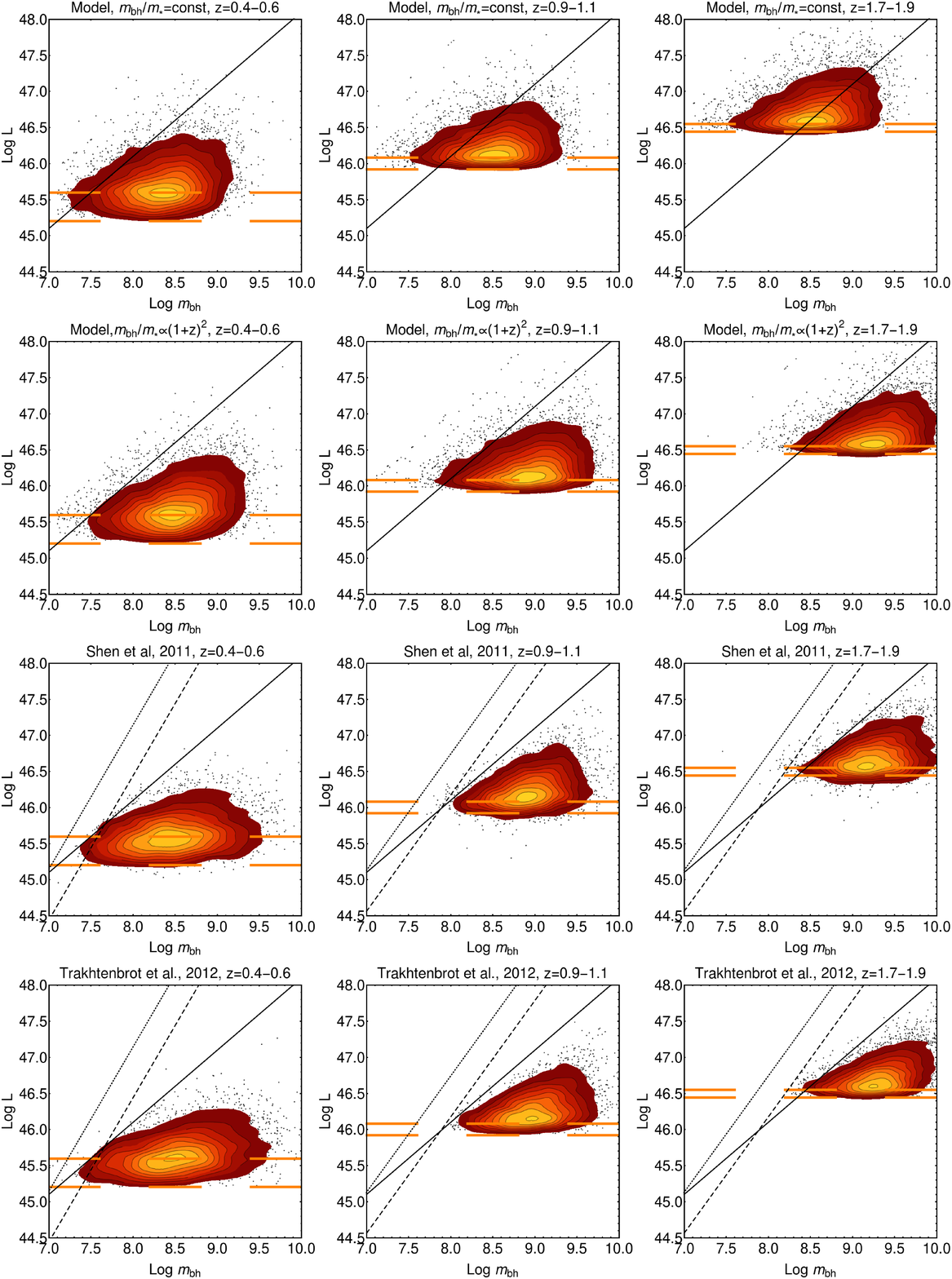}
	\caption{Mass-luminosity plane shown in 3 representative redshift bins. Our model with assumed non-evolving  relation $m_{bh}/m_{*}=10^{-2.8}$ is in the top row and our model with assumed evolving relation $m_{bh}/m_{*}=10^{-3}(1+z)^{2}$ is in the row below. Observational data from \cite{She11} and \cite{Tra12} are shown in bottom two rows. The thick black line is the Eddington limit, dashed orange lines show the calculated luminosity selection limit for lowest and highest redshift in the bin. The dotted and dashed black lines represent FWHM=1000 km/s and 1500 km/s, respectively, and are shown here to indicate which objects could be missed in observations because of FWHM limit in quasar selection. Contours are set at 10$\%$, 20$\%$ etc. values of estimed probability distribution of objects. Outermost objects are represented as individual dots.  } 	\label{fig:Lm1p2z0}
\end{figure*}

In this part of the paper we will compare predictions of how the black hole mass-luminosity plane of broad-line AGNs should be populated in our model, and compare these with SDSS data. After creating a mock sample of star-forming galaxies in an SDSS volume, we will populate them with black holes assuming different redshift-dependent $m_{bh}/m_{*}$ scaling relations and an assumed scatter.  We will then assign Eddington ratios from the evolving $\xi(\lambda,z)$ distribution and apply an obscuration prescription from \cite{Hop07}.  These two functions are chosen so that the QLF is reproduced as in the previous section: in other words the redshift evolution in $m_{bh}/m_{*}$ and the redshift evolution in $\lambda^*$ must combine to produce the observed evolution in the QLF $L^*$.   \\

For the observational distribution, we use data from two observational studies (\citealp{She11}; \citealp{Tra12}) to show that our results do not depend critically on the data choice and to give the reader a graphical impression of the uncertainties involved in this kind of measurement. By comparing our mock data with the observed distribution we will be able to see which combination of redshift-dependent $m_{bh}/m_{*}$ and $\xi(\lambda,z)$ best reproduces the observational data.  We take into account the obscuration factor and the differences in bolometric correction between different works and apply the same bolometric correction to the data and model (namely one used in \citealp{She11}) to make them directly comparable. \\

By using this mock sample approach we can fully account for biasses introduced by the luminosity-selection of the quasar samples.  We recreate data only up to $z = 2$, as black hole mass estimates for higher redshifts are based on the broad C IV emission line, which was shown to be far less reliable for these purposes (\citealp{She08}; \citealp{Tra12} and references therein). \\

We first show in the topmost panels of the Figure \ref{fig:Lm1p2z0} the modelled distribution of quasars in the black hole mass-luminosity $(m_{bh}, L)$ plane in 3 representative redshift bins if we assume a redshift independent $m_{bh}/m_{*}$ with the standard value of $m_{bh}/m_{*} \approx 10^{-2.8}$. We introduce a log-normal scatter in this relationship of 0.5 dex to account for scatter in $m_{bh}/m_{*}$ relationship.  The solid diagonal line indicates the Eddington limit ($\lambda = 1$).  \\

The data from (\citealp{She11}; \citealp{Tra12}) are plotted in the two bottom rows of panels in Figure \ref{fig:Lm1p2z0}.  In these panels, the diagonal dashed and dotted lines indicate the locus of black hole masses for two constant FWHM  (of 1000 km/s and 1500 km/s respectively) of the emission lines (H$\beta$ and MgII) that were used to infer the black hole masses.  Systems of lower FWHM will not appear in the samples, e.g. the limit used in \cite{She11} is set at 1200 km/s.  \\

We see that, although there is good agreement at low redshifts, a non-evolving $m_{bh}/m_{*}$ relation produces far too many objects with masses that are too small or, equivalently, which have very high Eddington ratios, with around 50$\%$ being super-Eddington in the final redshift bin, while only around 2$\%$ of objects are super-Eddington in the data. This is a simple consequence of the fact that $L^{*}$ is much higher then locally and is a reflection of the high $\lambda^*$ implied for an unevolving $m_{bh}/m_{*}$ relation shown in Figure \ref{fig:g21pz2}.\\

It should be noted that the problem gets worse if we reduce the scatter in the $m_{bh}/m_{*}$ relation as we then have even fewer high mass black holes.  We note that the comparison could not be expected to be perfect because our data is constrained to reproduce the QLF from \cite{Hop07}; although SDSS data and the optical QLF is the main contributor to the Hopkins QLF in this luminosity range, there are small contributions from other surveys as well as slightly different bolometric and obscuration corrections which will induce small differences. Nevertheless, the disagreement with a non-evolving $m_{bh}/m_{*}$ ratio is too large to be due to this.\\

A much better agreement is obtained (second row of Figure \ref{fig:Lm1p2z0}) if we adopt an evolving $m_{bh}/m_{*}$ relation.  We adopt as a heuristic example the form $m_{bh}/m_{*} \propto (1+z)^{n}$ with $n=2$. The agreement with the observed distribution is considerably better and there are now far fewer objects crossing the Eddington limit ($\sim$ 3 $ \%$) at high redshifts. This means that the observed $\sim (1+z)^4$ increase in $L*$ back to $z \sim 2$ would be due to an equal split between a $(1+z)^2$ change in $m_{bh}/m_{*}$ and a $(1+z)^2$ change in characteristic Eddington ratio $\lambda^*$, remembering that these two changes are degenerate in our convolution model without the $m_{bh}/m_{*}$ data of Figure \ref{fig:Lm1p2z0}.\\

Finally we note that, quite independently of any assumptions about $m_{bh}/m_{*}$, our convolution model naturally recreates the apparent ``sub-Eddington boundary'' that has been emphasized by \cite{Ste10a}, by which we mean the flat upper envelope . This refers to the fact that at all redshifts 
there seems to be a lack of objects at high masses close to the Eddington limit, which can also be seen in the Figure 16 of \cite{Tra12}. This can be observed on the Figure \ref{fig:Lm1p2z0}  where the upper contours of the red regions have slopes that are noticeably shallower than the 45 degree slope that corresponds to a constant Eddington ratio, thereby giving the impression of an absence of high luminosity high $\lambda$ sources.  \\

This behaviour is quite counter-intuitive, and was interpreted by \cite{Ste10a} as being caused by some new physical effect that somehow limits accretion onto more luminous quasars. Various authors have proposed alterations to the measurement methods in mass-luminosity plane which could reduce or eliminate this effect (e.g. \citealp{Raf11};  \citealp{Raf11b}). We show instead that this behaviour is actually expected from the convolution model presented in this paper! \\

We commented earlier in Section \ref{sec:Pred} that while the typical black hole mass increases with luminosity in our convolution model, it does so sub-linearly, so that the ``typical'' Eddington ratio must also increase with luminosity.  This can be seen also in these plots: the ridge line that is defined by the peak in the mass distribution \textit{ at a given $L$} is indeed {\it steeper} than the 45 degree line of constant Eddington ratio, which is why the {\it shallower} slope of the boundary defined by the contours is so counter-intuitive.  However, because the {\it number} of more massive black holes falls very steeply with mass, because of the decline of the galactic mass function above $M^*$, the contours of {\it constant surface density}, given by the contours in the plots in Figure \ref{fig:Lm1p2z0} and in the \cite{Ste10a} analysis, are actually {\it shallower} than the 45 degree line. \\ 

\begin{figure*}
	\centering
  \includegraphics[width=\textwidth]{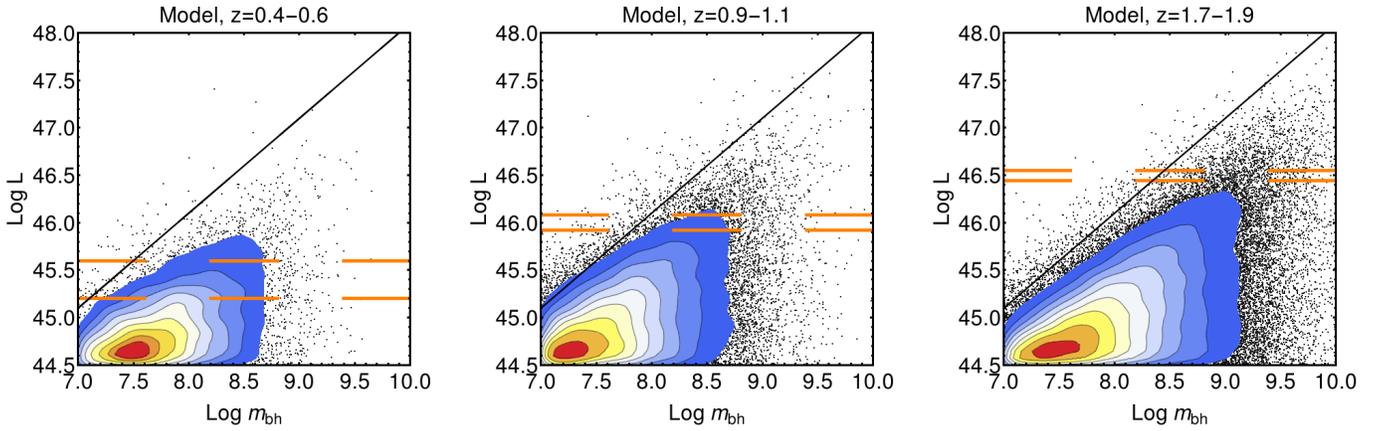}
	\caption{Full sample from our model, before applying the luminosity cuts to produce the second row in Figure \ref{fig:Lm1p2z0}. For clarity, only one out of every 25 points is shown.  } 
	\label{fig:Lm1p2z2Full}
\end{figure*}

This is more clearly seen in Figure \ref{fig:Lm1p2z2Full} were we show the appearance of our full sample, before the application of the SDSS luminosity cut. At lower luminosities the effect gets more and more pronounced and we expect that deeper surveys of a given area would find more super-Eddington objects at small masses. \\

We emphasize that the reproduction of the \cite{Ste10a} effect in Figure \ref{fig:Lm1p2z0} is achieved within our convolution model with an Eddington ratio distribution $\xi(\lambda)$ that is completely \textit{independent} of black hole mass.  The effect noted by \cite{Ste10a} would not be seen (in our model) if the distribution of points in the $(m_{bh}, L)$ plane was normalized to the total number of black holes (in star-forming galaxies) at a given \textit{mass}, which is, of course, unfortunately not observable. Our explanation for the sub-Eddington boundary in \cite{Ste10a} is therefore a kind of ``plotting bias'' arising from what is plotted, rather than a ``selection effect'' per se, coming through the construction of the sample via, e.g., emission line widths. 
\\ 

\subsection{Establishing the $m_{bh}/m_{*}$ correlation in quenched systems}\label{sec:quench}

In this section we show how it is possible to reproduce the observed tight correlation between $m_{bh}/m_{bulge}$ in the local Universe even if the black hole to stellar mass relation in the star-forming galaxies is strongly evolving (see also \citealp{Cro06}; \citealp{Hop0602}).  In what follows, we loosely equate $m_{bulge}$ with the stellar mass of a star-forming galaxy when star-formation quenches, which is also the point at which we assume the black hole ceases growing. Readers uncomfortable with making this association may simply substitute $m_{bulge}$ in what follows with $m_{quench}$.\\

\begin{figure*}
	\centering
  \includegraphics[width=\textwidth]{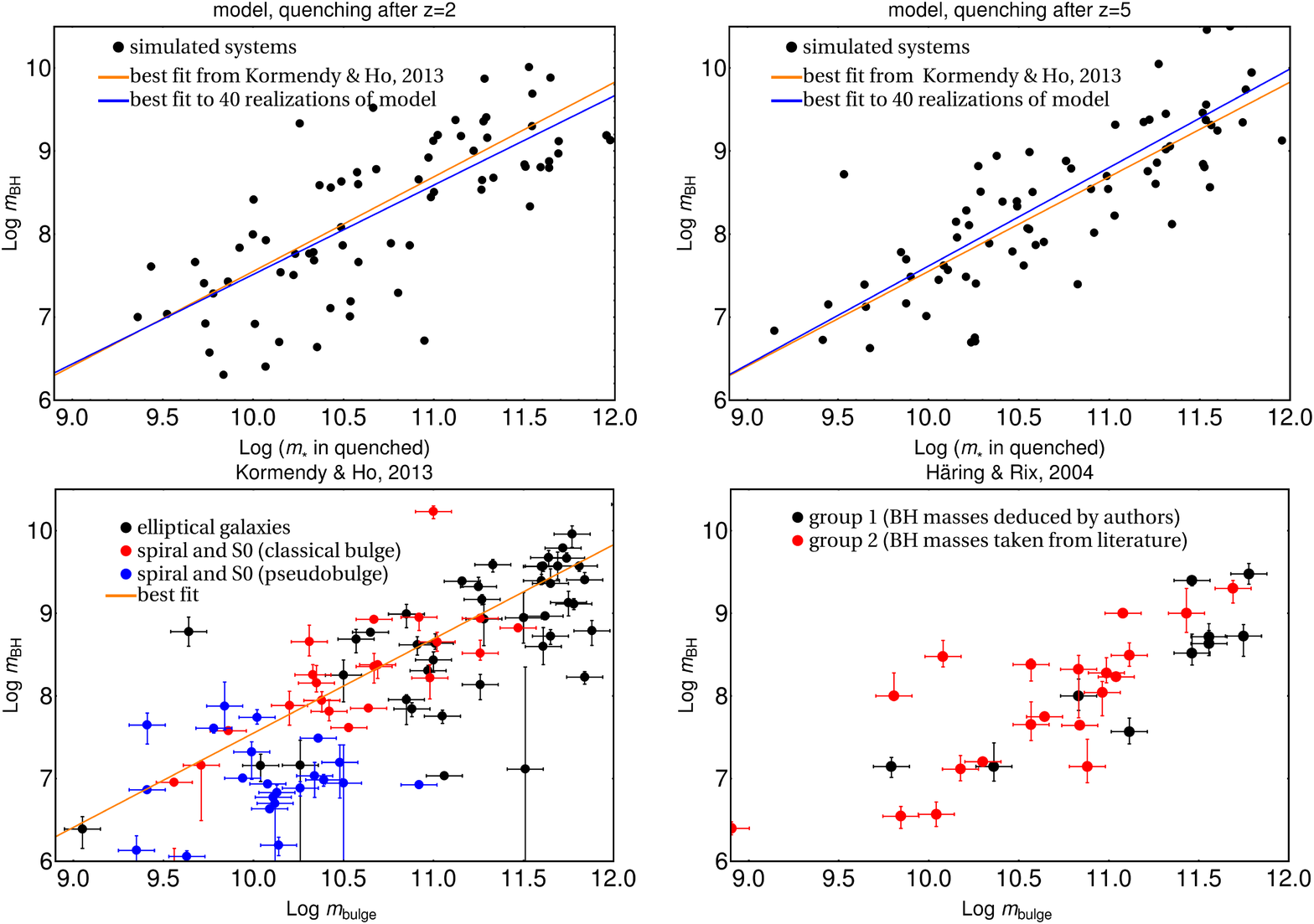}
	\caption{\textit{Top left}: Resulting $m_{bh}/m_{*}$ relation in quenched galaxy systems today, if  we only take into account systems that have quenched after $z=2$. Orange line shows best fit quoted in \cite{Kor13}. Blue line shows best fit to the data generated from 40 realizations of the model. \textit{Top right}: Resulting $m_{bh}/m_{*}$ relation in quenched systems today, for full range of quenching history ($z<5$). Orange and blue line are generated in same way as in previous panel. Merging, uncertain evolution in $m_{bh}/m_{*}$ relation and errors in estimated rate of quenching at high redshift could bias this result, as discussed in text. \textit{Bottom left}: Measured $m_{bh}/m_{bulge}$ by \cite{Kor13}. Orange line shows again best fit quoted in \cite{Kor13}. \textit{Bottom right}: Measured $m_{bh}/m_{bulge}$ by \cite{Har04}.}  
	\label{fig:Korm}
\end{figure*}
For this analysis, we use results from the simple galaxy evolution model of \cite{Bir14} to construct a set of evolving star-forming galaxies and their quenched descendants. The details of the \cite{Bir14} model are not important for the present purpose since it reproduces well the overall evolution of the galaxy population, which is all that we require here.   \\

For each quenched galaxy seen in the model at the present epoch, we know the mass and redshift at which it quenched and can therefore compute the black hole mass from the adopted redshift-dependent $m_{bh}/m_{*}$ relation for (star-forming) galaxies, adding also the adopted scatter. We assume that there is no stellar mass growth after quenching and that there is no central SMBH mass growth after quenching \\

After this procedure we are left with a mock sample of quenched galaxies, with their black hole masses known, after which we can account for sample selection effects. This is not trivial, as measurements of masses of black holes are from heterogeneous sources and no single luminosity or other cut is possible. \\

We therefore decided to create an empirical selection in galaxy mass so that the distribution of galaxy masses in the mock sample broadly matches that of the passive early type galaxies which have had their SMBH mass measured.  We show results for two situations, where we first only consider galaxies that quenched after $z=2$ and then include all galaxies that have quenched since $z=5$.
We differentiate between these two cases since we expect merging, which is not explicitly accounted for here, to have a larger impact for galaxies that quench earlier, and because the assumed mass ratio scaling, which we think is reasonable approximation (see above) at redshift $z<2$, may break down at higher redshifts. When we ignore quenching at $2 < z < 5$, we are losing only about 10\% of today's quenched population, but the model is on a much firmer footing. \\

We have fitted the simulated data, derived from 40 random realizations of the model,  with a relation of the form 
\begin{equation}
100 \left( \frac{ m_{bh}}{ m_{*, quench}} \right) = a \cdot \left( \frac{m_{*, quench}}{10^{11} M_{sun}} \right)^{b}
\end{equation}
by regressing the black hole mass on the stellar mass, and compute the scatter of the simulated galaxies around this relation.  We derive values of $(a,b)=(0.40,0.09)$ and a scatter of 0.53 dex for the case of quenching only from z=2, and values of $(a,b)=(0.63,0.18)$ with scatter of 0.6 dex for the case of quenching from z=5. These values should be compared with the observed values of $(a,b)=(0.49^{+0.06}_{-0.05},0.14 \pm 0.08)$ and a intrinsic scatter of 0.29 dex derived in \cite{Kor13}.   While the predicted scatter appears to be slightly larger than observed, subsequent merging of galaxies, that has not been modelled here, will act to reduce the scatter (\citealp{Hir10}, \citealp{Jah11}).\\

 The origin of $m_{bh}/m_{bulge}$ relation is the underlying $m_{bh}/m_{*}$ relation in star-forming galaxies assumed in the model, with added scatter due to the fact that galaxies of a given stellar mass today have quenched at various redshifts and, for a given stellar mass, the spread in black hole masses is amplified by spread in the quenching redshifts.  The mean of the $m_{bh}/m_{bulge}$relation is positioned roughly at the value of $m_{bh}/m_{*}$ at the mean redshift of quenching at that mass, which is $z \sim 1-1.5$
over a wide range of galaxy masses. \\

It is interesting to note that objects that have quenched more recently would be expected on average to have a lower $m_{bh}/m_{*}$ value, because of the evolution in $m_{bh}/m_{*}$.  It is possible that these recently quenched galaxies could be associated with pseudo-bulges instead of bulges.  \cite{Kor13} has indeed argued that pseudo-bulges do indeed have a lower $m_{bh}/m_{bulge}$ ratio. \\

We also note in passing that if there is a trend for the typical quenching redshift to increase with increasing stellar mass (e.g. because of interplay of mass and environment quenching, \citealp{Pen10}) then this would introduce a tilt in the local $m_{bh}/m_{bulge}$ relation ($b \neq 0$) even though the input $m_{bh}/m_{*}$ in star-forming galaxies was perfectly linear.  \\

The link with quenching redshift then prompts another interesting point.  There is very good evidence that the sizes of galaxies, at a given mass, are smaller at high redshift (\citealp{Dad05};  \citealp{New12};  \citealp{Car13};  \citealp{Sha13b}). We would therefore expect, through virial arguments, that the velocity dispersions, at a given mass, would also be higher.  This has been directly observed in a few cases (e.g. \citealp{VanD14}).  For analytic simplicity, we assume that the scale radius of galaxies scales as $(1+z)$ so that the usual Faber-Jackson type scaling relation would be expected to evolve as

\begin{equation}
m_{star} \propto \sigma^4 (1+z)^2. 
\end{equation} 
If $m_{bh}/m_{*}$ scales as $(1+z)^2$, as we have been exploring in this part of the paper, then this naturally produces an $m_{bh}-\sigma$ relation that is \textit{independent} of redshift.  \\

This has two implications: first, we would not expect to see any significant evolution in the observed $m_{bh}-\sigma$ relation (see e.g. recently \citealp{She15}).   Second, we would expect
the present-day $m_{bh}-\sigma$ relation for passive galaxies to be tighter than the $m_{bh}-m_{*}$ relation, because of the aforementioned broadening of the latter from the range of $z_{quench}$. \\

We stress that this tighter $m_{bh}-\sigma$ relation would be present even if the velocity dispersion is not playing a direct role in the growth of black holes.  Of course, it is also possible that the evolution in the  $m_{bh}/m_{*}$ is in fact \textit{caused} by the constancy of an  underlying $m_{bh}-\sigma$ causal connection.  On the other hand, there could well be other causes for $m_{bh}/m_{*}$ to evolve as $(1+z)^2$, in which case the tight $m_{bh}-\sigma$ relation would simply be a coincidence.\\

\subsection{$m_{bh}/m_{*}$ in AGNs in the local Universe} \label{subsec:Mat}

Finally, we turn to estimates of the $m_{bh}/m_{*}$  relations for AGNs in the local Universe, where it is possible to estimate independently the mass of the galaxy that is hosting an optically active AGN. \cite{Mat14} made a careful decomposition into nuclear and host contributions of the images of $z< 0.6$ quasars in the SDSS Stripe 82. They found that the quasars are predominately hosted in massive star-forming galaxies, with relatively large $m_{bh}/m_{*}$ ratios of around $10^{-2.5}$. \cite{Mat14} were aware of the possibility that selection effects could bias this value upwards but could not estimate their magnitude.  We can now use our model to examine the expected size of this bias.\\

\begin{figure}
	\centering
  \includegraphics[width=0.5\textwidth]{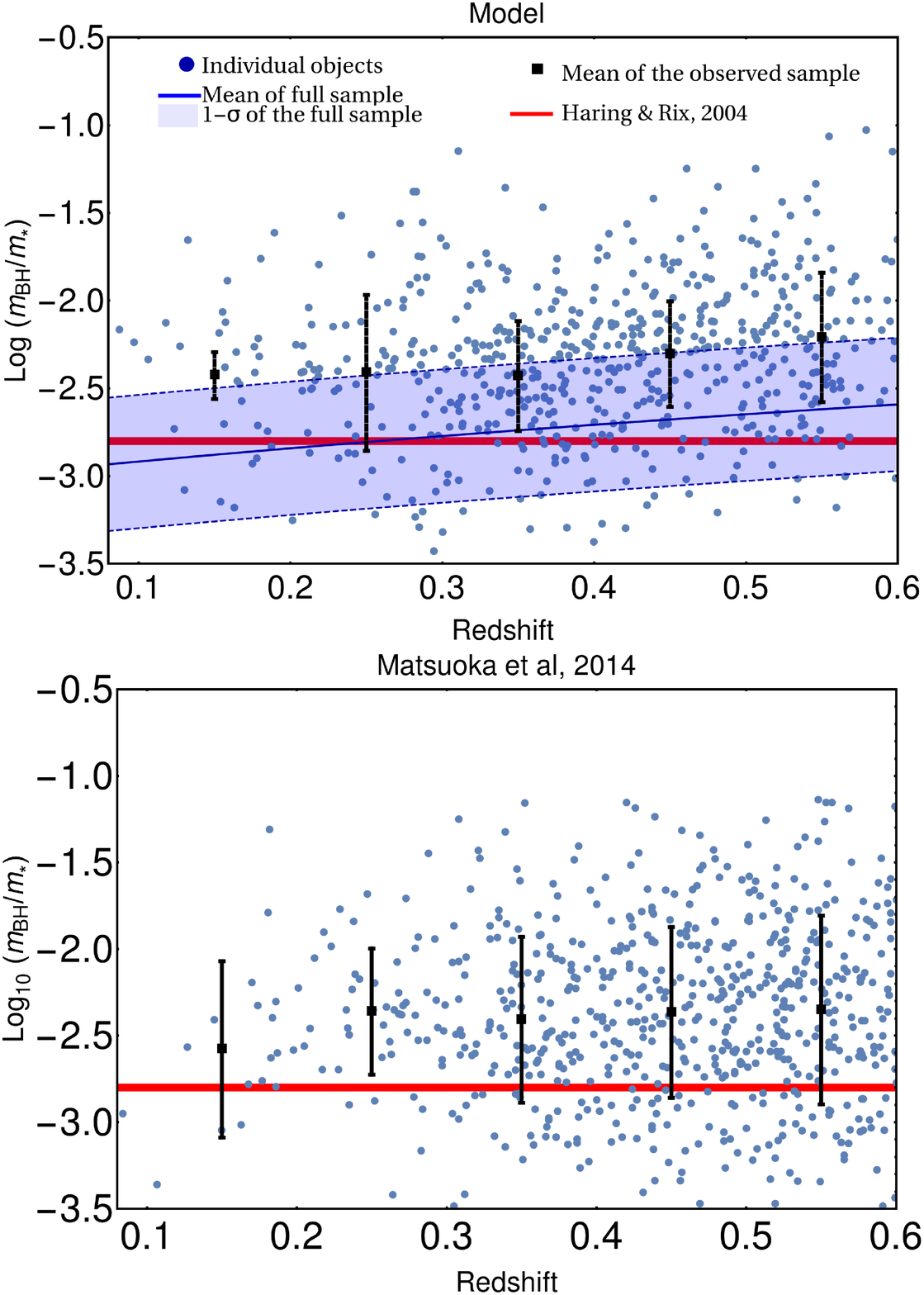}
	\caption{\textit{ Top panel}: Simulated observation of $m_{bh}/m_{*}$ relation in Stripe 82, mimicking analysis of \cite{Mat14}. Blue line shows mean $m_{bh}/m_{*}$ relation in full sample (before luminosity cut was applied), while the shaded area shows $1-\sigma$ spread around mean relation in our model. The black line shows mean and spread of distribution of points in 0.1 redshift slices. Red line is set to $10^{-2.8}$, which is approximately local relation from \cite{Har04}. \textit{Bottom panel}: original data from \cite{Mat14}. } 
	\label{fig:Mat}
\end{figure}

To do this we simulate sources within the same sky area as Stripe 82 and recreate objects above the AGN luminosity cut that would have been selected for quasar spectroscopy in the SDSS sample, with scatter of $\eta=0.4$ dex. The results are shown in Figure \ref{fig:Mat}, represented in the same way as in the original paper.  We see that the observed quasars will have $m_{bh}/m_{*}$ ratios that are indeed much higher then the mean value in the underlying sample, which is indicated by the shaded region in the figure. \\

Finally, we notice that even though we inserted $m_{bh}/m_{*} \propto (1+z)^2$ redshift evolution in our model, this evolution would be quite hard to detect in the sample of \cite{Mat14}, owing to the large spread of points and the selection biasses connected with such a study. Nevertheless, we do still see a slight change in the mean values of the simulated sample that is not seen in the actual data. We discuss this further in Section \ref{subsec:LowRedMstar} below.\\

\section{Discussion} \label{sec:Dis}

In the preceding Sections of this paper we have developed a simple generic model for obtaining the evolving AGN luminosity function from a convolution of the evolving galaxy mass function.  This generic model makes testable predictions for quantities such as the shape of the mass distribution of host galaxies as a function of AGN luminosity and allows us to derive quantitative connections between the parameters describing the galaxy mass function and the AGN QLF.  On the basis of these, we can derive powerful statements about the duty cycle of AGN.  We then showed, in the framework of this general model, that a redshift-dependent $m_{bh}/m_{*}$ and Eddington ratio distribution, $\xi(\lambda)$, successfully reproduces the observed quasar luminosity function (by construction) and also reproduces observations of the distribution of quasars in the ($m_{bh}, L$) plane, the black hole to bulge mass relation of quenched galaxies and measurements of $m_{bh}/m_{*}$ for low redshift AGN.  In this Discussion section, we develop further some of the astrophysical implications of the preceding results.

\subsection{Comparing $m_{bh}/m_{*}$ redshift evolution in active and quenched systems} 

It is important to appreciate that a quite rapid evolution in $m_{bh}/m_{*}$ in \textit{active} (star-forming) AGN systems does not imply an equally rapid evolution of $m_{bh}/m_{*}$ in the \textit{quenched} systems. This is because, when we are observing quenched systems, we are effectively observing the integrated history of previous activity. To illustrate this, we perform a simple heuristic exercise in which we determine $m_{bh}/m_{*}$ in quenched systems at given epoch, assuming as above that quenching started at redshift 2. We make the same assumptions as before, namely that there is no stellar or black hole mass growth after quenching. \\

\begin{figure}
	\centering
  \includegraphics[width=80mm]{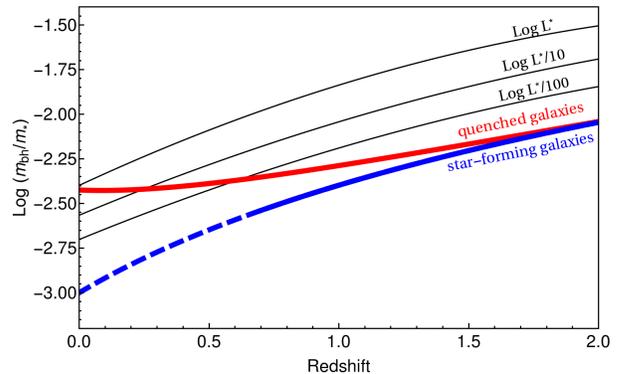}
	\caption{ Evolution of the mean $m_{bh}/m_{*}$ relation in star-forming (heavy blue line) and quenched systems (heavy red line). The dashed blue line shows the redshift range where this mean relation may not accurately represent star-forming galaxies, as discussed in Section \ref{subsec:LowRedMstar}. The evolution in $m_{bh}/m_{*}$ for the quenched (inactive nuclei) galaxies is always shallower than for the population of actively accreting systems, potentially leading to an underestimate of the evolution if the populations are mixed. On the other hand, the ``Lauer effect'' will bias the $m_{bh}/m_{*}$ upwards in luminosity selected active samples, which goes in the opposite direction.  This is indicated by the thin black lines which show the observed $m_{bh}/m_{*}$ for active systems for luminosity-selection (labelled relative to $L^*$) for our standard assumption of $\sigma$ = 0.5 dex.}
	\label{fig:Bul}
\end{figure}

Our results are shown in Figure \ref{fig:Bul}. Even though star-forming galaxies have changed their black-hole to stellar mass ratio by almost a factor of 10 from redshift $z \sim 2$ to today, the evolution in this ratio for \textit{quenched} systems is much milder, more like a factor of 3 or even less (0.4 dex), simply because the quenched population includes galaxies that have quenched much earlier.  This means that the redshift evolution in $m_{bh}/m_{*}$ for quenched systems will always be much milder than the evolution in this quantity in active systems.\\

The distinction between active and passive populations becomes even more important if observational studies compare actively accreting systems at high redshift with data on the quiescent population at low redshift.   This is unfortunately quite common practice (e.g. \citealp{Jah09}; \citealp{Dec10}; \citealp{Mer10}; \citealp{Ben11}; \citealp{Schr13}; \citealp{Schu15}), because of the current practicalities of observations. We stress that this approach automatically produces weaker evolution since neither $m_{bh}$ nor $m_{*}$ will have changed once a given object becomes inactive (i.e. passive/quenched).  It is therefore clear that the mean $m_{bh}/m_{*}$  relation in the quenched systems will reflect the mass ratio in the star-forming galaxies at the much earlier epochs when those galaxies actually quenched (as already discussed above).  For the local population, this is typically around z $\sim$ 1-1.5.\\

Clearly, if we compare the $m_{bh}/m_{*}$  of star forming systems at z $\sim$ 1-1.5 to the $m_{bh}/m_{*}$ ratio of passive galaxies seen today that quenched at z $\sim$ 1-1.5, then we would expect to see no change in $m_{bh}/m_{*}$, even if this ratio had changed a lot within the actively accreting population! Of course, if the high redshift active systems are luminosity-selected then their $m_{bh}/m_{*}$ ratio will likely have been biassed to higher values than the underlying population through the ``Lauer effect'' discussed in Section 3.4, which goes in the opposite direction (as shown by the parallel black lines in Figure \ref{fig:Bul}).  This figure shows that if Nature has an underlying $m_{bh}/m_{*}$ scaling as $(1+z)^2$ for active systems, then we would expect to see $(1+z)^{0.8}$ if we compare comparisons of active systems at $z \sim 2$ with quenched systems at $z \sim 0$, once we correct for these observational biases.   This is quite similar to the evolution seen in at least some of the observational studies cited above (e.g. \citealp{Jah09};  \citealp{Dec10}; \citealp{Mer10}; \citealp{Schr13}; \citealp{Schu15}).  \\

We have already commented in Section \ref{sec:quench} that we would also expect to see little or no evolution in $m_{bh} - \sigma$ for any population of galaxies, however selected. This is due to the higher $\sigma$ associated with a given stellar mass at high redshift which cancels out the $(1+z)^2$ dependence in $m_{bh}/m_{*}$.

\subsection{Mass growth of star-forming galaxies at low redshift }\label{subsec:LowRedMstar}

In this paper we have made the conventional assumption that the black hole to stellar mass relation increases as some power of (1+z), i.e. $m_{bh}/m_{*} \sim (1+z)^{n}$.  However, this is an arbitrary redshift dependence, and there are reasons why it cannot hold (with $n \sim 2$) at low redshifts (see also \citealp{Hop06}). Assuming that a star-forming galaxy is on the Main Sequence we can track the increase of its stellar mass using the Main Sequence sSFR(z),

\begin{equation} \label{eq:rsSFR}
rsSFR (z)=\frac{\dot{m}_{*}}{m_{*}} =  \frac{-1}{(1+z)\sqrt{\Omega_{M}(1+z)^{3}+\Omega_{\Lambda}}} \frac{1}{m_{*}} \frac{d m_{*}}{dz},
\end{equation}

where $rsSFR$ is the ``reduced specific star-formation rate'' (see \citealp{Lil13}), $rsSFR = (1-R)sSFR$, where R is the fraction $R$ of stellar mass that is returned during star formation $R \sim 0.4$, which is therefore the inverse mass doubling timescale of the stellar population. Using $rsSFR (z) = 0.07 (1+z)^{3}$ Gyr$^{-1}$ from \cite{Lil13} and references therein, this can be integrated to give 

\begin{equation}\begin{split}
m_{*}(z)= m_{*}(0)\left( \exp\left[  \frac{2 \cdot 0.07 \sqrt{\Omega_{M}+\Omega_{\Lambda}}}{3 H_{0} \Omega_{M}}  \right. \right. \\ 
\left. \left. - \frac{2 \cdot 0.07 \sqrt{\Omega_{M}(1+z)^{3}+\Omega_{\Lambda}}}{3 H_{0} \Omega_{M}} \right]  \right). 
\end{split}\end{equation}

\begin{figure}
	\centering
  \includegraphics[width=80mm]{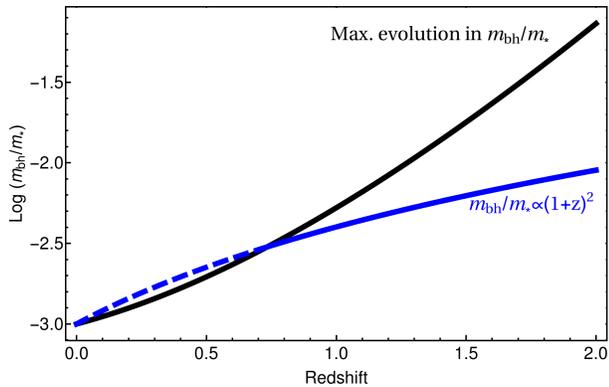}
	\caption{Comparison of the maximal evolution of the $m_{bh}/m_{*}$ relation in a star-forming galaxy that stays on the Main Sequence, with our assumed $m_{bh}/m_{*} \propto (1+z)^{2}$. At low redshift galaxies are growing so slowly in the stellar mass that strong evolution in the $m_{bh}/m_{*}$ ratio is not possible, even if the black hole does not grow at all.  } 
	\label{fig:MaxEv}
\end{figure}

This modest increase in stellar mass for a star-forming galaxy sets a maximal change in $m_{bh}/m_{*}$ ratio even in the most extreme case that $m_{bh}$ does not increase at all.  We show this maximal evolution in Figure \ref{fig:MaxEv} and compare it with the $(1+z)^2$ evolution used elsewhere in the paper. It can be seen that the maximal evolution is actually slower then $(1+z)^{2}$ at redshifts $z \lesssim 0.7$.   Galaxies are growing so slowly at low redshifts that it is not possible to create a strong evolution in $m_{bh}/m_{*}$ ratio in a given galaxy, even if their black holes are not growing at all. This may well be why there is no evidence in Figure \ref{fig:Mat} for the "expected" increase in the observed $m_{bh}/m_{*}$.  The maximal change in $m_{bh}/m_{*}$ over this redshift range would have been gentle enough that it would be very hard to observe.  For this reason, we emphasize that our statements elsewhere in the paper concerning the evolution in $m_{bh}/m_{*} \propto (1+z)^2$ should best be interpreted as implying a change of a factor of about ten to $z \sim 2$, rather than a precise particular dependence on redshift.

\subsection{Downsizing}

A number of authors (e.g. \citealp{Has05}; \citealp{Bar05}; \citealp{Lab09}; \citealp{Li11}) have noted or discussed a ``downsizing'' of the quasar population. Although different authors often use this term to mean different things, it is most often used to denote the observational fact that lower-luminosity AGNs peak in comoving density at lower redshifts then higher-luminosity AGNs. \\

\begin{figure}
	\centering
  \includegraphics[width=80mm]{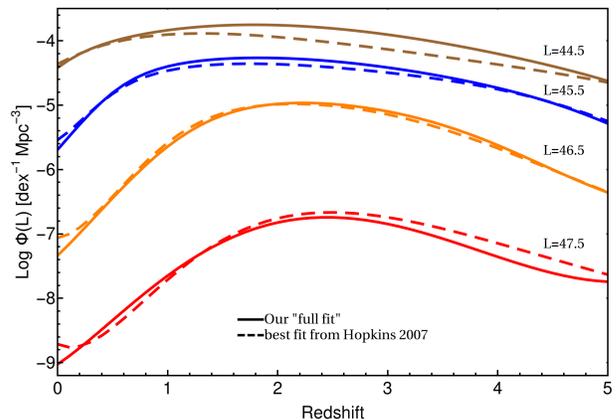}
	\caption{Redshift evolution of comoving density of quasars for several different luminosities. The density of lower luminosity AGN peaks at lower redshift then for high luminosity AGNs. This behaviour is naturally produced in our model even though the distribution of Eddington ratios is completly independent of black hole mass. }  
	\label{fig:Hasinger}
\end{figure}

It is worth stressing that this may not have much physical significance.  We have shown that it is possible to reproduce the strong observed redshift evolution in the QLF with a model based on the observed mass function of star-forming galaxies coupled with a \textit{mass-independent} (but redshift-dependent) Eddington ratio distribution $\xi(\lambda,z)$.   This is shown more explicitly in Figure \ref{fig:Hasinger} where we show the comoving number density of quasars of different luminosity in the QLF which is reproduced by our model.  This emphasizes that the apparent down-sizing signature in the AGN population can appear even though the distribution of Eddington ratios (and thus of specific black hole growth rates) is strictly {\it independent} of black hole mass at all redshifts in our model.\\

It is clear that the apparent ``downsizing'' in our model arises as a natural consequence of two competing effects which are independent of mass.  The first is the redshift evolution of the $L^*$ which shifts the luminosity function {\it uniformly} in luminosity, but which therefore produces a differential change in number density with luminosity. This shift is produced by the degenerate combination of evolution of the $m_{bh}/m_{*}$ mass ratio and characteristic Eddington ratio, $\lambda^{*}$. The second is the redshift evolution of $\phi^*_{QLF}$ that changes the number densities {\it uniformly} with luminosity. We have argued in this work that the evolution of $\phi^*_{QLF}$ is a direct consequence of the evolution of $\phi^*_{SF}$, coupled with a constant duty cycle.  The combination of these two produces variation in the number density (at fixed luminosity) that changes with luminosity, producing a variation in the redshift of the peak in the number density as well as different rates of evolution at different luminosities.

\subsection{Coherent evolution of $\phi^{*}_{SF}$ and $\phi^{*}_{QLF}$}

One of the most striking results of this paper is that the observed evolution of $\phi^{*}_{QLF}$ of the QLF tracks the observed evolution of $\phi^{*}_{SF}$ of the star-forming galaxy mass function.  This in turn implies that the $\xi^*$ knee of the Eddington ratio distribution $\xi(\lambda)$ has a more or less constant value, even though the change of the characteristic $\lambda^*$ "knee" is dramatic.  We have argued that $\xi^*$ is a good measure of a generalized ``duty cycle'' of quasars. Indeed, if the luminosity of individual quasars decays with a timescale $\tau$ then (Equation \ref{fd}) $\xi^*_{\lambda}$ will be direct measure of the birth-rate of quasars in star-forming galaxies. \\

By applying simple continuity equations to the observed evolution of the star-forming galaxy mass function, \cite{Pen10} derived an expression for the rate of the mass-quenching process, $\eta_{m}$, which may be written

\begin{equation}\label{quench}
\eta_{m} \sim sSFR(z) \cdot (m_{*}/M^*),
\end{equation}

where sSFR(z) is the redshift dependent specific star-formation rate of the star-forming Main Sequence.  It has been well established that sSFR was much higher in the past and a useful representation is (\citealp{Lil13} and references therein)

\begin{equation} \label{eq:sSFR}
sSFR \sim  \begin{cases}
 (1+z)^{3}, & \text{when }z< 2 \\
 (1+z)^{5/3}, & \text{when }z > 2.
 \end{cases}
\end{equation}

AGN activity in "quasar-mode" is one of the many processes that have been proposed (e.g. \citealp{Gra04}; \citealp{Hop08}; \citealp{Kin10}) to drive the mass-quenching of galaxies.   If a single quasar event is responsible for quenching, then this would require that $\eta_{AGN}$ (from Equation \ref{fd}) to be equal to $\eta_{m}$. These can only be equated for our inferred constant $\xi^*_{\lambda}$ for a particular redshift and mass dependence of the decay time $\tau$.

\begin{equation}\label{equate}
\tau(m,z) \sim sSFR(z)^{-1} (m/M^*)^{-1} \xi^*_{\lambda}.    
\end{equation}

We would require quasars to fade faster at high redshift and at high galaxy masses.  We could well imagine ways in which this would occur, e.g. because of the shorter dynamical timescales of galaxies at high redshift.\\

However, the above picture is probably over-simplistic. We could well expect the physics of quenching to be more complex. It is quite plausible that the energy source for quenching is AGN activity but that the effectiveness of this energy injection depends on the stellar or halo mass of the system (the \citealp{Pen10} quenching law can be written in terms of a redshift-independent survival probability that depends only on mass).   This would then break the simple link between $\eta_{m}$ and $\eta_{AGN}$.    \\

\section{Summary and conclusions} \label{sec:SaC}

We have presented a simple, phenomenological model that aims to link the evolving galaxy population with the evolving AGN population. We use our observational knowledge of the evolving galaxy mass function and of the evolving quasar luminosity function (QLF) to connect these two populations and to create a global model to interpret the AGN population, including biases associated with sample selection. \\

Our model is based on three observationally motivated Ans\"atze, namely that 
\begin{itemize}
   \setlength{\itemsep}{1pt}
  \setlength{\parskip}{0pt}
  \setlength{\parsep}{0pt}
\item { radiatively efficient AGNs are found in star-forming galaxies,  }
\item { the probability distribution of the Eddington ratio does not depend on the black hole mass of the system,  }
\item { the mass of the central black hole is linked to the stellar mass.}
\end{itemize} 

\noindent  The QLF is then a straightforward convolution of the black hole mass function with the distribution of Eddington ratios $\xi(\lambda,z)$, while the former is itself a convolution of the galaxy mass-function with the $m_{bh}/m_*$ relation.  
These heuristic assumption ensure that our model is simple enough to be analytically tractable, while still capturing the main characteristics of the galaxy and AGN population. \\

The main conclusions of this model analysis can be summarized as follows:

\begin{enumerate}

\item 
The ``broken'' or ``double" power law form of the quasar luminosity function is a consequence of the underlying Schechter mass function and a ``double" power law, mass independent, Eddington ratio distribution. We show how the parameters of the QLF are straightforwardly connected with the input functions. Most importantly, the knee of the QLF, $L^{*}$, is proportional to the product of the $M^{*}$ of the galaxy mass function, the ratio $m_{bh}/m_{*}$ and the position $\lambda^{*}$ of the break in the Eddington ratio distribution while the $\phi^{*}_{QLF}$ normalization of the QLF is proportional to the product of the $\phi^{*}_{SF}$ normalization of the star-forming galaxy mass function and the $\xi^{*}_{\lambda}$ normalisation of the Eddington ratio distribution, which can be loosely interpreted as a "duty cycle". \\

Our simple convolution model makes clear and testable predictions for the distribution of host galaxy masses (relative to the star-forming galaxy Schechter $M^*$) for different AGN luminosities (relative to $L^*$).  At high luminosities (above the AGN $L^*$) this is a Schechter function with the star-forming M* but a faint end slope given by $\alpha_{SF}+\gamma_2 \sim 1.5$.

\item 
There is a remarkable consistency in the redshift evolution of $\phi^{*}_{SF}$ normalization of SF mass function and the $\phi^{*}_{QLF}$ normalization of QLF.  These two characteristic densities track each other closely out to redshifts of $z \sim 3$, and possibly to even higher redshifts.  This implies that the generalised ``duty cycle'' of AGN is surprisingly constant with redshift.

\item 

In contrast, the QLF $L^{*}$ evolves strongly with redshift, with evolution being at least $L^{*} \propto (1+z)^{3}$ up to $z \sim 2$. Given that there is strong evidence for no change in the galaxy $M^{*}$ over this redshift range, this evolution in $L^*$ is driven by an evolution in the characteristic ``knee'' in the Eddington ratio distribution $\lambda^{*}$ or in the mass scaling between stellar mass and black hole mass, $m_{bh}/m_{*}$, or some combination of the two.  The QLF evolution is degenerate in changes of these two quantities.

\end{enumerate}

We then explore this degeneracy by comparing predictions of our model, incorporating the relevant selection cuts, for the distribution of systems in the SDSS AGN mass-luminosity plane(s) using black hole mass estimates for individual AGN.   We find a good match with an evolving  $m_{bh}/m_{*} \propto (1+z)^{2}$ in star-forming systems.  This implies that the observed $\sim (1+z)^4$ increase in $L*$ back to $z \sim 2$ would be due to an equal split between a $(1+z)^2$ change in $m_{bh}/m_{*}$ and a $(1+z)^2$ change in characteristic Eddington ratio $\lambda^*$. \\

We show that this is compatible with the observed $m_{bh}/m_*$ relations in both quenched and in star-forming galaxies in the local Universe, both in terms of the mean relations and the scatter. \\

We also make the important point that much weaker evolution, more like $m_{bh}/m_{*} \propto (1+z)^{0.8}$, would be deduced  by observers, after correcting for the ``Lauer bias'', if (as they usually of necessity do) they compare black hole masses in active (star-forming) systems at high redshift ($z \sim 2$) with those in quiescent systems at low redshift.   Similarly, much weaker evolution would also be seen if black hole masses were to be compared solely within the passive population at different redshifts.\\

The inferred $m_{bh}/m_{*} \propto (1+z)^{2}$ evolution in star-forming systems is likely simplistic, and unlikely to hold all the way down to the lowest redshifts simply because there is not enough star-formation to change $m_{*}$ fast enough, even if black holes are not growing at all.   Our use of a $(1+z)^{2}$ evolutionary model should be interpreted more loosely as a way of getting a factor of 2.5 change between $m_{bh}/m_{*}$ in {\it active} systems at $z \sim 2$ when compared with the same relation in {\it passive} systems today, rather than as a precise relation that absolutely holds at all redshifts. \\

We however make the interesting point that an evolution of the form $m_{bh}/m_{*} \propto (1+z)^{2}$ would, when coupled with the observed changes in the sizes of galaxies, which typically scale as roughly $(1+z)^{-1}$, have the feature of producing a $m_{bh} - \sigma$ relation that would be more or less \textit{independent} of redshift.  This could then explain why the present-day $m_{bh} - \sigma$ relation would have lower scatter than the $m_{bh}/m_{*}$ relation, even if $\sigma$ plays no direct role in black hole growth.    Alternatively, a fundamental redshift-independent $m_{bh} - \sigma$ relation could be the physical origin of an apparent $m_{bh}/m_{*} \propto (1+z)^{2}$ evolution.\\

We stress that the most basic features of the model do not depend on the redshift dependence of $m_{bh}/m_{*} $, which is driven largely by the possibly uncertain observational estimates of black hole masses in high redshift AGN. \\

Not least, quite independent of the form of any evolution in $m_{bh}/m_{*}$, the generic model naturally reproduces the counter-intuitive ``sub-Eddington boundary'' in the ($m_{bh}-L$) plane that has been noted by \cite{Ste10a} without the need to invoke any new physical effects.   The generic model also produces the apparent ``down-sizing'' of the AGN population (e.g. \citealp{Has05}).  We stress that both of these apparently mass-dependent effects are achieved in the model with an Eddington ratio distribution that is completely \textit{ independent} of black hole mass at all redshifts.\\

{\bf Acknowledgements}  We thank Simon Birrer for providing us with the full output of the model presented in \cite{Bir14}. We also thank Peter Behroozi for providing us with the galaxy mass functions that were used as input in \cite{Ber13}. Sebastian Seehars has helped in creating fitting procedures. In developing this work, we have benefited from stimulating discussions with Charles Steinhardt, Andreas Schulze, Kevin Schawinski and David Rosario.  We are also very grateful for the very perceptive comments provided by the anonymous referee.  This work has been supported by the Swiss National Science Foundation.

\renewcommand{\theequation}{A-\arabic{equation}}
  
  \setcounter{equation}{0}  
  \section*{APPENDIX A:}

Here we will show analytically how the convolution of a Schechter mass function $\phi_{BH}(m_{bh})$ with a triangular Eddington ratio distribution $\xi(\lambda)$ in Equation (\ref{Ed}) gives rise to a double power-law QLF and the simple relations between parameters given in (\ref{eq:Master}). In this simplest case when $\xi(\lambda) = 0$ below $\lambda^*$, combining equations (\ref{eq:CreateQLF}),  (\ref{Ed}) and (\ref{eq:SF}) leads to

\begin{equation}\begin{split} \label{eq:ap1}
\phi(L) & = \int^{\infty}_{\lambda^{*}} \phi^{*}_{SF} \xi^{*}_{\lambda} \left( \frac{L}{10^{38.1}  M^{*}_{BH} \lambda  } \right)^{\alpha_{BH}} \exp\left[ \frac{-L}{10^{38.1} M^{*}_{BH} \lambda }\right] \left( \frac{\lambda}{\lambda^{*}}\right)^{-\delta_{2}} \frac{ 1}{\ln(10)\lambda }  d \lambda \\
& = \frac{  \phi^{*}_{SF} \xi^{*}_{\lambda}}{\ln(10)} \left(\frac{10^{38.1} M^{*}_{BH}\lambda^{*}}{L}\right)^{\delta_{2}} \left( \Gamma[\alpha_{BH} + \delta_{2}] - \Gamma[\alpha_{BH} + \delta_{2}, \frac{L}{10^{38.1} M^{*}_{BH}\lambda^{*}}\right) ,
\end{split} \end{equation}
where $\Gamma[x]$ is Euler gamma function, given by $\Gamma[x]=\int^{\infty}_{0} t^{x-1} e^{-t}dt$ and $\Gamma[x,y]$ is incomplete gamma function, defined as $\Gamma[x,y]=\int^{\infty}_{y} t^{x-1} e^{-t}dt$.   
On the other hand we know that the integral of distribution of Eddington ratios has to be equal to $f_d$
\begin{equation}
\int^{\infty}_{0} \frac{\xi (\lambda)}{\log (10) \lambda} d \lambda = 1  \rightarrow  \int^{\infty}_{\lambda^{*}} \xi^{*}_{\lambda}\left( \frac{\lambda}{\lambda^{*}}\right)^{-\delta_{2}} \frac{d \lambda}{\ln(10)\lambda}=f_{d},
\end{equation}
which gives
\begin{equation} \label{eq:ap3}
\xi^{*}_{\lambda}= f_{d}\delta_{2}\ln(10).
\end{equation}
\begin{figure}
	\centering
  \includegraphics[width=0.7 \textwidth]{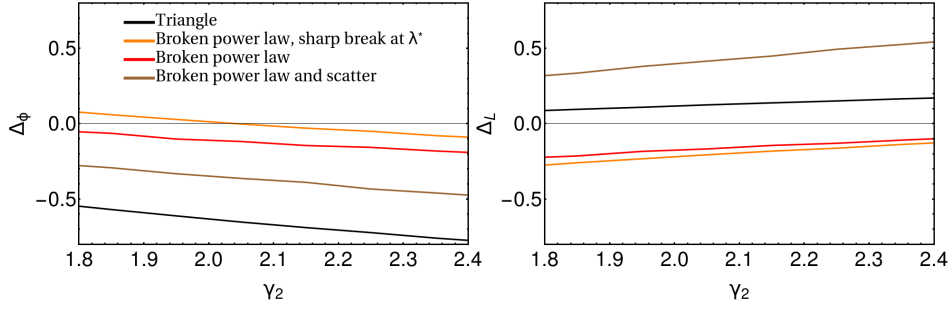} 
	\caption{Functional dependence of the factors $\Delta_L(\delta_{1},\gamma_2)$ and $\Delta_\phi(\delta_{1},\gamma_2)$. Black line corresponds to the "triangle" distribution given with Equation (\ref{Ed}). Orange line is showing the result for the distribution shown in Equation (\ref{eq:sharp}) (with $\delta_{1}=$0), red line is showing the result for the broken power law distribution, given in Equation (\ref{Ed2}) (with $\delta_{1}=0$), while brown line corresponds to previous case, but with added scatter when converting from the star-forming galaxy mass function to the mass function of black holes in star-forming galaxies. } 
	\label{fig:DoubePowerLawConv2}
\end{figure}

Combining Equations (\ref{eq:ap1}) and (\ref{eq:ap3}) we arrive at an expression for the QLF

\begin{equation} \label{eq:ap4}
\phi (L)= f_{d} \phi^{*}_{SF} \cdot \left\lbrace  \delta_{2}  \left( \frac{10^{38.1} M^{*}_{BH} \lambda^{*}}{L} \right)^{\delta_{2}}  \left( \Gamma[\alpha_{BH} + \delta_{2}] - \Gamma[\alpha_{BH} + \delta_{2}, \frac{L}{10^{38.1} M^{*}_{BH}\lambda^{*}}]\right) \right\rbrace ,
\end{equation}

where we separate the "normalization" from the rest of the expression in curly brackets.  We can expand this expression at low and high $L$ to show asymptotic power law behaviour. Expanding  around $L=0$ gives
\begin{equation}
\phi (L \rightarrow 0)= f_{d}\phi^{*}_{SF}\delta_{2}\cdot L^{\alpha_{BH}} \cdot \left(  \frac{(10^{38.1}M^{*}\lambda^{*})^{-\alpha_{BH}}}{\alpha_{BH}+\delta_{2}} - \frac{(10^{38.1}M^{*}\lambda^{*})^{-\alpha_{BH}}}{10^{38.1}M^{*}_{BH}\lambda^{*}(1+\alpha_{BH}+\delta_{2})}L +\mathcal{O}(L^{2}) \right), 
\end{equation}
so we see that dominant term will be $L^{\alpha_{BH}}$ and that $\gamma_{1}=-\alpha_{BH}$. This is special case of our formula $\gamma_{1}= \mbox{max}(-\alpha_{BH},\delta_{1})$ as in this case $\delta_{1}$ is effectively minus infinity. Expanding around $L \rightarrow \infty$,
\begin{equation}
\phi (L \rightarrow \infty)= f_{d}\phi^{*}_{SF}\delta_{2}\cdot \left( \frac{10^{38.1} M^{*}_{BH}\lambda^{*}}{L}\right)^{\delta_{2}} \cdot \left( \Gamma[\alpha_{BH}+\delta_{2}]-\exp\left[-\frac{L}{10^{38.1} M^{*}_{BH}\lambda^{*}} \right] \left( \frac{L}{10^{38.1}M^{*}\lambda^{*}} \right)^{-1+\alpha_{BH}+\delta_{2}}   +\mathcal{O}\left( \frac{1}{L^{2}}\right) \right) ,
\end{equation}
we see that dominant term is $L^{-\delta_{2}}$, giving rise to the equation $\gamma_{2}=\delta_{2}$.\\
Defining the double power-law as in Equation (\ref{eq:QLFF}), we find  expression for $L^{*}$ and $\phi^{*}$ is given by

\begin{equation} \begin{split} \label{eq:ap6}
L^* & \approx 10^{38.1} M^{*}_{BH}\lambda^{*} \\
\phi^{*}_{QLF} &  \approx \phi^{*}_{SF}  f_{d}\cdot \delta_{2} \left( \Gamma[\alpha_{BH}+\delta_{2}]-\Gamma[\alpha_{BH}+\delta_{2},1]\right) \\
& \approx  \phi^{*}_{SF}  f_{d}.
\end{split}\end{equation}

As would be expected the $L^{*}$ and $\phi_{QLF}^{*}$ are related to the input $M^{*}_{BH}$ and $\phi_{SF}^{*}$ via the characteristic $\lambda^{*}$ and the normalization of $\xi^{*}_{\lambda}$ given by $f_d$. \\
We have numerically fitted the resulting $\phi(L)$ around $L^*$ with the broken power law given in equation (\ref{eq:QLFF}) (with $\gamma_1 =- \alpha_{SF}$ ) for various $\gamma_{2}$ values and plot the resulting offsets $\Delta_L$ and $\Delta_{\phi}$ in Figure \ref{fig:DoubePowerLawConv2} such that 
\begin{figure}
	\centering.
  \includegraphics[width=80mm]{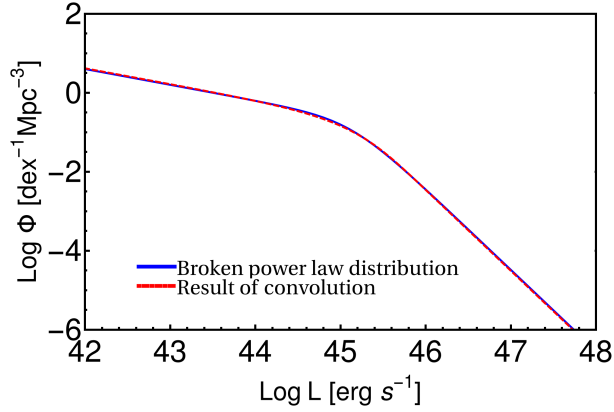} 
	\caption{ QLF in a shape of a broken power law distribution is show in blue, while in red we show results of convolution of Schechter mass function and simple "triangle" distribution of Eddington ratio (Equation (\ref{eq:ap4})), with parameters set to reproduce blue line. As we see, agreement is excellent.  } 
	\label{fig:DoubePowerLawConv}
\end{figure}

\begin{equation} \begin{split}\label{eq:ap7}
L^* &= 10^{38.1} \cdot \lambda^{*} \cdot M^{*}_{BH} \cdot  \Delta_L(\gamma_2) \\
\phi^{*}_{QLF}& =\phi^{*}_{SF} \cdot \xi^{*}_{\lambda}\cdot \Delta_\phi(\gamma_2)
\end{split}\end{equation}
It can be seen that, as would be expected from Equation (\ref{eq:ap4}), the values of $\Delta_L$ and $\Delta_{\phi}$ vary with $\gamma_2$.
The standard double power-law Equation (\ref{eq:QLFF}) is a surprisingly good representation of the analytic result in Equation (\ref{eq:ap4}). Example of power-law shape and result from our convolution is shown in Figure \ref{fig:DoubePowerLawConv}. \\
We can also repeat this exercise with a second form of $\xi_{\lambda}$ that is flat below $\lambda^*$, i.e. has $\delta_1 = 0$, replacing $f_d$ by $\xi^{*}_{\lambda}$, i.e. 
\begin{equation} \begin{split}\label{eq:ap8}
L^* &= 10^{38.1} \cdot \lambda^{*} \cdot M^{*}_{BH} \cdot  \Delta_L(\delta_{1},\gamma_2) \\
\phi^{*}_{QLF}& =\phi^{*}_{SF} \cdot \xi^{*}_{\lambda}\cdot \Delta_\phi(\delta_{1},\gamma_2)
\end{split}\end{equation}
where we incorporate the fact that $\Delta$ factors depend on the assumed shape of the Eddington ratio. This can again be seen in Figure \ref{fig:DoubePowerLawConv2}, where we also show results of fitting with Eddington ratio distribution which has sharp break at $\lambda^{*}$, i.e.
\begin{equation} \label{eq:sharp}
\xi (\lambda) =  \begin{cases} \xi^{*}_{\lambda}\left( \frac{\lambda}{\lambda^{*}} \right)^{-\delta_{1}} & \lambda < \lambda^{*}\\ 
\xi^{*}_{\lambda}\left( \frac{\lambda}{\lambda^{*}} \right)^{-\delta_{2}} & \lambda > \lambda^{*}. \end{cases}
\end{equation}

  \section*{APPENDIX B:} 
  
\begin{figure*}
	\centering
  \includegraphics[width=.99\textwidth]{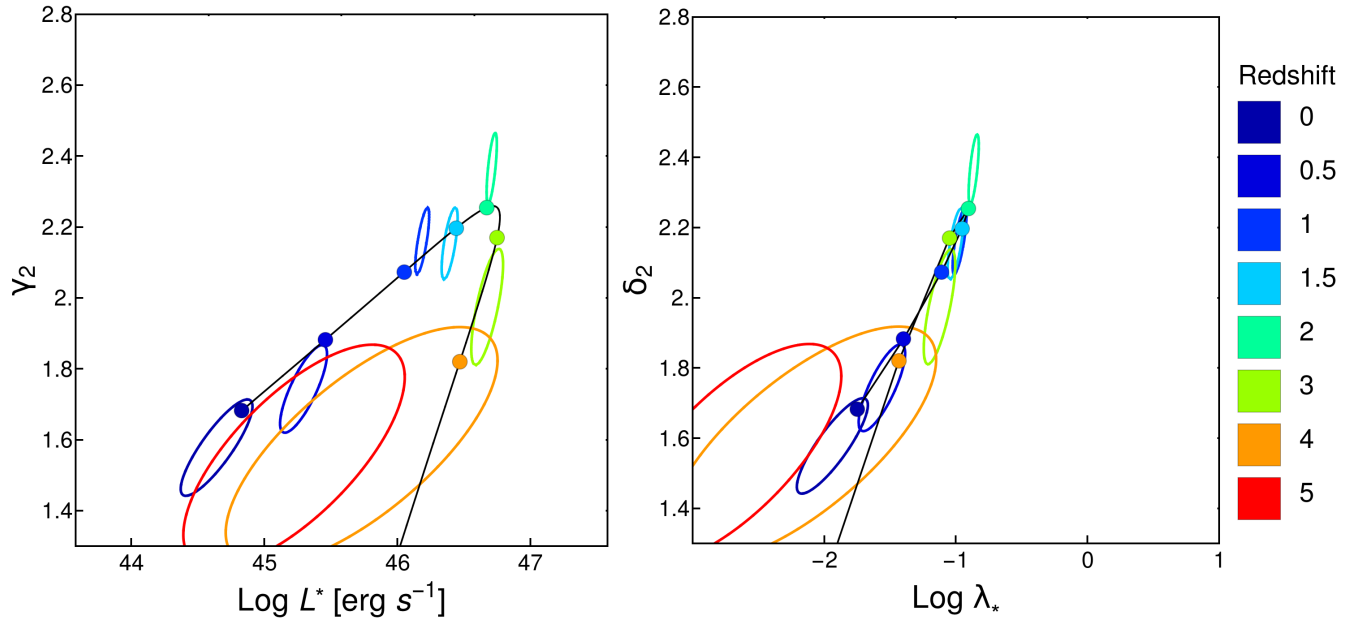}
	\caption{\textit{Left:} $\gamma_{2}-L^{*}$ redshift evolution. As in Figure \ref{fig:ParEvoFull}, contours are showing 1-$\sigma$ allowed regions of parameter space for fit which was made  with data at each individual redshift, while filled circles are result of global fit. \textit{Right:} $\delta_{2}-\lambda^{*}$  evolution derived from $\gamma_{2}-L^{*}$ evolution  with our knowledge of redshift evolution of $M^{*}$ and with assumed mass scaling $m_{bh}/m_{*} \propto (1+z)^{2}$. All of the points and contours seem to align along single line suggesting intimate link between these two quantities.}
	\label{fig:g21pz2}
\end{figure*}

In this section we wish to point out one interesting and suggestive fact that arises from fits that were discussed in Section \ref{sec:Comparing}.
We show on the left hand side of Figure \ref{fig:g21pz2} the relation obtained between $L^{*}$ and the bright end slope of the QLF, $\gamma_{2}$. There is a clear increase in the bright end slope $\gamma_{2}$ as $L^{*}$ increases with redshift.  At $z \geq 3$, as $ L^{*}$ stalls and then declines, we also see $\gamma_{2}$ decreasing again, although not in unison with $L^{*}$ in the sense that the track in the ($L^{*},\gamma_{2}$) plane is displaced. We can plot an equivalent diagram in the ($\lambda^{*},\delta_{2}$) plane by converting $ L^{*}$ to $\lambda^{*}$ using an $m_{bh}/m_*$ ratio and Equation (\ref{eq:MasterL}) and setting $\delta_{2} = \gamma_2$ (Equation (\ref{eq:gamma2})).  The steepening of $\delta_2$ as $\lambda^{*}$ increases
might qualitatively be expected if there was some maximum value of the Eddington ratio $\lambda_{max}$.  As $\lambda^*$ increased towards such a limit, then $\xi(\lambda)$ above $\lambda^*$ would have to steepen, so as to get down to zero at the limiting $\lambda_{max}$. 
 
An interesting question is then whether we can find an evolving (i.e. redshift-dependent) $m_{bh}/m_*$ ratio that will cause the ``up'' and ``down'' tracks in Figure \ref{fig:g21pz2} to lie on top of each other.  We find that introducing $m_{bh}/m_{*}\propto (1+z)^{2}$ produces a good congruence in the ($\lambda^{*},\delta_{2}$) tracks associated with the rise and decline of $L^*$.   This is shown in the right hand panel, where we show the ($\lambda^{*},\delta_{2}$) tracks obtained by using $m_{bh}/m_{*} = 10^{-3} (1+z)^{2}$.  The congruence with this particular $m_{bh}/m_{*}$ evolution is suggestive but cannot be taken as a strong indication of this particular evolution of $m_{bh}/m_{*}.$\\

However, the values of $\lambda^*$ that are implied (in our convolution model) at high redshifts by the high $L^*$ are getting uncomfortably high, unless there is a strong increase in $m_{bh}/m_{*}$ relative to the canonical zero redshift value of $10^{-2.8}$,  As can be seen on the right hand plot of Figure \ref{fig:g21pz2}. With $(1+z)^{2}$ evolution, $\lambda^*$ is implied to be -0.8 dex at $z \sim 2$, but without this evolution, $\lambda^*$ is implied to be +0.15 dex, i.e. above the Eddington limit! This would mean that the characteristic Eddington ratio at this redshift would be above Eddington limit, with many of the objects being super-Eddington. Both observations and theoretical work suggests that the fraction of AGNs which cross this limit is small (e.g. \citealp{Woo02}; \citealp{Tra07}; \citealp{Hop10}; \citealp{She11}; \citealp{Tra12}).  \\  

\end {document}